\documentclass[iop,apj]{emulateapj}
\usepackage{apjfonts}

\begin{document}

\title{GRB 130427A and SN 2013cq: A Multi-wavelength Analysis of An Induced Gravitational Collapse Event}

\author{R.~Ruffini\altaffilmark{1,2,3,4}, Y.~Wang\altaffilmark{1,2}, M.~Enderli\altaffilmark{1,3}, M.~Muccino\altaffilmark{1,2}, M.~Kovacevic\altaffilmark{1,3}, C.L.~Bianco\altaffilmark{1,2}, A.V.~Penacchioni\altaffilmark{4}, G.B.~Pisani\altaffilmark{1,3}, J.A.~Rueda\altaffilmark{1,2,4}}

\altaffiltext{1}{Dip. di Fisica and ICRA, Sapienza Universit\`a di Roma, Piazzale Aldo Moro 5, I-00185 Rome, Italy.}
\altaffiltext{2}{ICRANet, Piazza della Repubblica 10, I-65122 Pescara, Italy.}
\altaffiltext{3}{Universit\'e de Nice Sophia Antipolis, CEDEX 2, Grand Ch\^ateau Parc Valrose, Nice, France.}
\altaffiltext{4}{ICRANet-Rio, Centro Brasileiro de Pesquisas Fisicas, Rua Dr. Xavier Sigaud 150, Rio de Janeiro, RJ, 22290-180, Brazil.}

\shorttitle{GRB 130427A and SN 2013cq}

\shortauthors{Ruffini et al.}

\begin{abstract}
We have performed our data analysis of the observations by \textit{Swift}, \textit{NuStar} and \textit{Fermi} satellites in order to probe the induced gravitational collapse (IGC) paradigm for GRBs associated with supernovae (SNe), in the ``terra incognita'' of GRB 130427A.
We compare and contrast our data analysis with those in the literature.
We have verified that the GRB 130427A conforms to the IGC paradigm by examining the power law behavior of the luminosity in the early $10^4$ s of the XRT observations. This has led to the identification of the four different episodes of the ``binary driven hypernovae'' (BdHNe) and to the prediction, on May 2, 2013, of the occurrence of SN 2013cq, duly observed in the optical band on May 13, 2013. The exceptional quality of the data has allowed the identification of novel features in \textit{Episode 3} including: a) the confirmation and the extension of the existence of the recently discovered ``nested structure'' in the late X-ray luminosity in GRB 130427A, as well as the identification of a spiky structure at $10^2$ s in the cosmological rest-frame of the source; b) a power law emission of the GeV luminosity light curve and its onset at the end of \textit{Episode 2}; c) different Lorentz $\Gamma$ factors for the emitting regions of the X-ray and GeV emissions in this \textit{Episode 3}.
These results make it possible to test the details of the physical and astrophysical regimes at work in the BdHNe: 1) a newly born neutron star and the supernova ejecta, originating in \textit{Episode 1}, 2) a newly formed black hole originating in \textit{Episode 2}, and 3) the possible interaction among these components, observable in the standard features of \textit{Episode 3}.

\end{abstract}


\email{yu.wang@icranet.org}

\maketitle

\date{}

\section{Introduction and Summary of Previous Results}\label{sec:1}

That some long gamma-ray bursts (GRBs) and supernovae (SNe) can occur almost simultaneously has been known for a long time, since the early observations of GRB 980425/SN 1998bw \citep{Galama:1998ea,Pian:2000dd}. This association of a GRB and a SN occurs most commonly in a family of less energetic long GRBs with the following characteristics: 1) isotropic energies $E_{iso}$ in the range of ${10^{49}}$--${10^{52}}$~erg \citep{Guetta:2007tb}; 2) a soft spectrum with rest-frame peak energy $E_{p,i} < 100$~keV, although the instruments are sensitive up to GeV; 3) supernova emissions are observable up to a cosmological distance $z<1$. We shall refer to this family in the following as \textit{family 1}. This result has been well recognized in the literature, see e.g. \citep{Maselli:2013hc}.

There is an alternative family of high energetic long GRBs possibly associated with SNe which have a much more complex structure. 
Their characteristics are: 1) $E_{iso}$ is in the range ${10^{52}}$--${10^{54}}$~erg; 2) they present multiple components in their spectra and in their overall luminosity distribution, ranging from X-ray, $\gamma$-ray all the way to GeV emission. They have peak energies from $100$~keV to some MeV; 3) in view of their large energetics, their observation extends to the entire universe all the way up to $z=8.2$ \citep{2014arXiv1404.1840R}. We shall refer to this family in the following as \textit{family 2}. 

Doubts were advanced that SNe may be associated with very bright long GRBs: naive energetic arguments considered that there can hardly be a SN in a powerful GRB within the single star collapse model \citep[see e.g.][]{Maselli:2013hc}.

The issue of the coincidence of very energetic GRBs with SN has represented for some years an authentic ``terra incognita''. The crucial point is to clarify whether this association of GRBs and SNe is only accidental or necessary, independent of their energetics. Up to June 2014, out of $104$ long GRBs with known redshift $z<1$, 19 GRBs associated with SNe belonging to the \textit{family 2} have been observed \citep{Kovacevic:2014up}, and GRB 130427A with isotropic energy $E_{iso}\simeq 10^{54}$~erg is the most energetic one so far.

\begin{figure}
\centering
\includegraphics[width=\hsize]{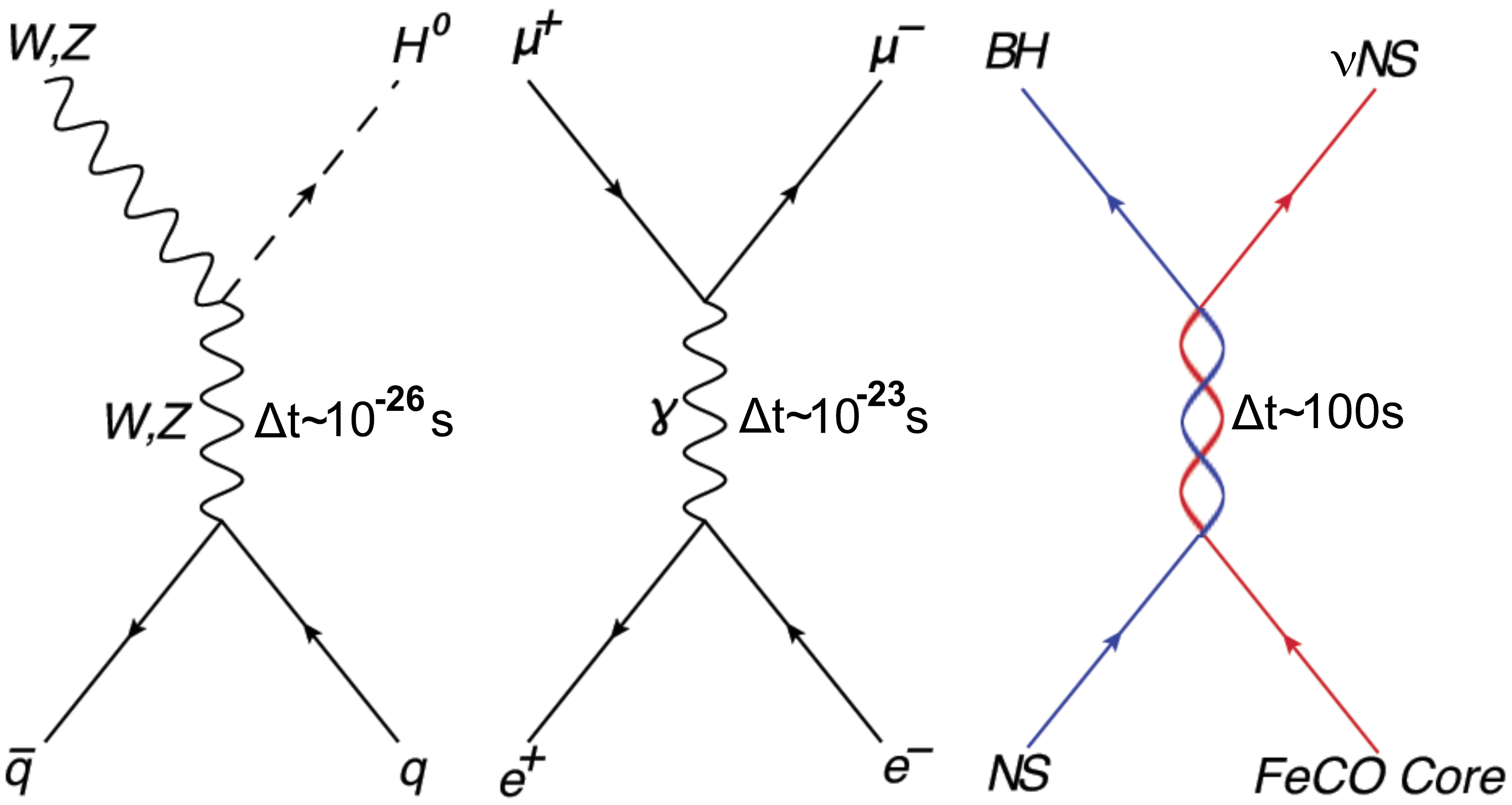}
\caption{Three different matrices in fundamental physics. The first is the quark matrix leading to a Higgs boson. In the middle is the classical electron-positron pair matrix, generating an muon and anti-muon pair. The third matrix is the most recent one, which is considered in the present work. $\Delta$t is the duration of intermediate state.}
\label{matrix}
\end{figure}

In \citep{2001ApJ...555L.117R,2008mgm..conf..368R}, we have introduced the paradigm of induced gravitational collapse (IGC) in order to explain the astrophysical reasons for the association of GRBs with supernovae.
This paradigm indicates that all long GRBs, by norm, have to be associated with SNe. The IGC paradigm differs from the traditional collapsar-fireball paradigm \citep[see, e.g.,][and references therein]{Piran:2005cs}.
In the collapsar-fireball model  the GRB process is described by a single episode: 1) it is assumed to originate in a ``collapsar'' \citep{1993ApJ...405..273W}; 2) the spectral and luminosity analysis is typically time integrated over the entire $T_{90}$ \citep[see e.g.][]{1998ApJ...497L..21T}; 3) the description of the afterglow is dominated by a single ultra-relativistic jetted emission \citep[see, e.g., in][]{1999ApJ...525..737R,vanEerten:2010gw,vanEerten:2012hp,Nava:2013gf}.
In contrast, the IGC paradigm considers a multi-component system, similar to the ones described by $S$-matrix in particle physics as shown in Figure \ref{matrix}: 1) the ``in-states'' are represented by a binary system formed by a FeCO core, very close to the onset of a SN event, and a tightly bound companion neutron star (NS) \citep{2008mgm..conf..368R,2012ApJ...758L...7R,2012A&A...548L...5I}. The ``out-states'' are the creation of a new NS ($\nu$-NS) and a black hole (BH). In the case of particle physics the $S$-matrix describes virtual phenomena occurring on time scales of $10^{-26}$~s \citep[][$q\bar{q} \rightarrow WZH^{0}$]{2012PhLB..716....1A} and $10^{-23}$~s \citep[][$e^{+}e^{-} \rightarrow \mu^{+}\mu^{-}$]{Bernardini:2004cl}. In the astrophysical case, here considered, the cosmic matrix ($C$-matrix) describes real event occurring on timescale $\sim200$~s, still a very short time when compared to traditional astrophysical time scales. Following the accretion of the SN ejecta onto the companion NS binary, a BH is expected to be created, giving origin to the GRB; 2) special attention is given to the analysis of the instantaneous spectra in optical, X-ray, $\gamma$-ray and GeV energy range (as exemplified in this article); 3) four different episodes are identifiable in the overall emission, each with marked differences in the values of their Lorentz $\Gamma$ factors \citep{2014A&A...565L..10R}. Actually the possible relevance of a binary system in the explanation of GRBs was already mentioned in a pioneering work of \citet{1999ApJ...526..152F} and in \citet{2005MNRAS.361..955B}, but the binaries in their case were a trigger to the traditional collapsar model.

The opportunity to probe the IGC paradigm \citep{2012A&A...548L...5I} has come from the prototypical source GRB 090618, a member of \textit{family 2}. This source, has an extremely high energetics, i.e., $E_{iso} = 2.7 \times 10^{53}$ erg, is at a relatively close distance, i.e., $z=0.54$, and has a coverage by all the existing $\gamma$, X-ray and optical observatories.

\begin{figure}
\centering
\includegraphics[width=\hsize]{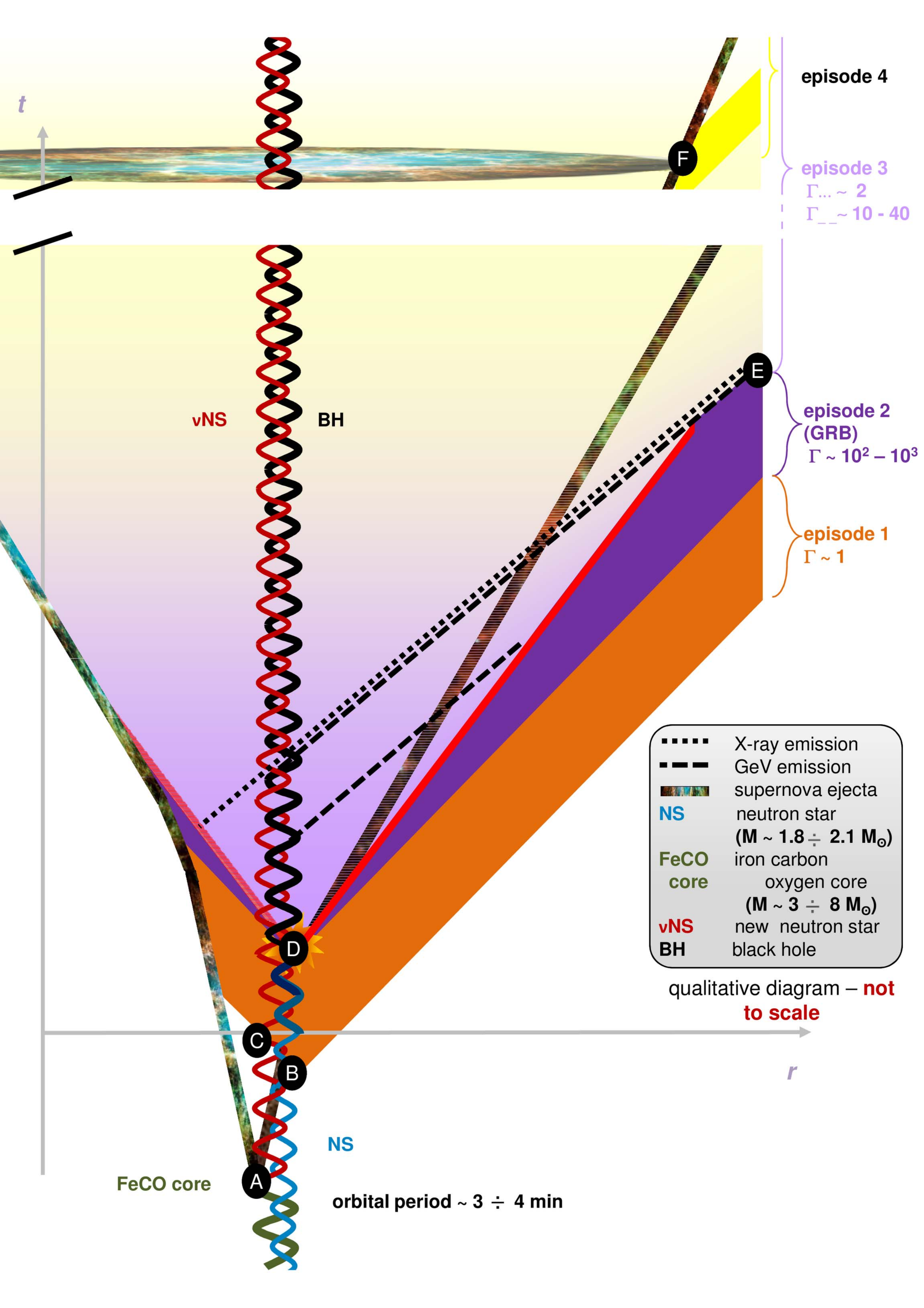}
\caption{IGC space–time diagram (not in scale) illustrates 4 episodes of IGC paradigm: the non-relativistic \textit{Episode 1} ($\Gamma \simeq 1$), the relativistic motion of \textit{Episode 2} ($\Gamma \simeq 10^{2} \sim 10^{3}$), the mildly relativistic \textit{Episode 3} ($\Gamma \simeq 2$), and non-relativistic \textit{Episode 4} ($\Gamma \simeq 1$). Initially there is a binary system composed by a massive star (yellow thick line) and a neutron star (blue line). The massive star evolves and explodes as a SN at point A, forms a $\nu$NS (red line). The companion NS accretes the supernova ejecta starting from point B, interacts with the $\nu$NS starting from point C, and collapses into a black hole  (black line) at point D, this period  from point B to point D we define as\textit{Episode 1}. Point D is the starting of \textit{Episode 2}, due to the collision of GRB outflow and interstellar filaments. At point E, \textit{Episode 2} ends and \textit{Episode 3} starts, \textit{Episode 3} lasts till the optical signal of supernova emerges at point F, where the \textit{Episode 4} starts. (\textit{Credit to M.Enderli on drawing this visualized space-time diagram.})
}
\label{spacetime}
\end{figure}

A wealth of results have been obtained:

1) \textit{Episode 1} corresponding to the onset of the SN and the accretion process onto the companion NS was soon identified in the early $50$~s, with a thermal plus power law component in the spectra \citep[see][Fig.~16]{2012A&A...548L...5I}, as well as a temporal evolution of the radius of the emitting region expanding  from $10^9$~cm to $7 \times 10^9$~cm \citep[see][Fig.~18]{2012A&A...548L...5I}, leading to a precise determination of its overall energetics of $4 \times 10^{52}$~erg.

2) \textit{Episode 2} with the GRB emission, following the onset of gravitational collapse and the BH formation, was also clearly identified with the characteristic parameters: an isotropic energy $E_{iso} = 2.49 \times 10^{53}$, baryon loading $B=1.98 \times10^{-3}$, Lorentz factor $\Gamma=495$ \citep[see][Fig.~4]{2012A&A...548L...5I}, and peak energy $E_{p,i} = 193$~keV. The average number density of the circumburst medium (CBM) is $\langle n_{CBM}\rangle=0.6$ cm$^{-3}$. The characteristic masses of each CBM cloud have been found to be of the order of $\sim10^{22}$--$10^{24}$ g, at $10^{16}$ cm in radii \citep[see][Fig.~10]{2012A&A...548L...5I}.

3) \textit{Episode 3} of GRB 090618, detected by \textit{Swift}-XRT, starts at $150$~s after the burst trigger and continues all the way up to $10^6$~s. It consists of three different parts \citep{2006ApJ...642..389N}: a) a first very steep decay; b) a shallower decay, the plateau and c) a final steeper decay with a fixed power law index. It soon became clear that this \textit{Episode 3}, which had been interpreted in the traditional approach as part of the GRB  afterglow \citep{Piran:2005cs,1999ApJ...525..737R,vanEerten:2010gw,vanEerten:2012hp,Nava:2013gf}, appeared to be the seat of a set of novel independent process occurring after the end of the GRB emission and preceding the optical observation of the SN, which we indicated as \textit{Episode 4}.

Recently, progress has been made in the analysis of \textit{Episode 1}. It is characterized by the explosion of the FeCO core, followed by the hypercritical accretion onto the NS which leads to the reaching of the critical mass of the NS and consequently to its induced gravitational collapse to a BH. The hypercritical accretion of the SN ejecta onto the NS has been estimated using the Bondi-Hoyle-Lyttleton formalism to be $10^{-2}~M_{\odot}~s^{-1}$, here $M_{\odot}$ is the solar mass \citep{Bondi:1944ty,1952MNRAS.112..195B}, see e.g., in\citep[][]{2012ApJ...758L...7R}. The inflowing material shocks as it piles up onto the NS, producing a compressed layer on top of the NS \citep[see e.g.,][]{1996ApJ...460..801F}. As this compressed layer becomes sufficiently hot, it triggers the emission of neutrinos which cool the in-falling material, allowing it to be accreted into the NS \citep{1972SvA....16..209Z,1975PhRvD..12.2959R,  1999A&A...350..334R,2000A&A...359..855R,1996ApJ...460..801F,1999ApJ...526..152F}. Recently \citet{2014arXiv1409.1473F} have presented a significant progress in understanding the underlying physical phenomena in the aforementioned hypercritical accretion process of the supernova ejecta into the binary companion neutron star \citep{2008mgm..conf..368R,2012ApJ...758L...7R}. The new treatment, based on the two-dimensional cylindrical geometry smooth particle hydrodynamics code, has simulated numerically the process of hypercritical accretion, the classical Bondi-Hoyle regimes, in the specific case of the IGC paradigm and leading to the first astrophysical application of the neutrino production process considered in \citet{1972SvA....16..209Z} and in \citet{1975PhRvD..12.2959R}, see e.g., in R. Ruffini et al. presentation in Zeldovich-100 meeting\footnote{http://www.icranet.org/index.php?option=com$\_$content$\&$task=view$\&$ \\ id=747$\&$Itemid=880}. Indeed the fundamental role of neutrinos emission allows the accretion rate process to increase the mass of the binary companion star to its critical value and lead to the black hole formation giving origin to the GRB in \textit{Episode 2}. This results confirm and quantifies the general considerations presented in\citep[][]{2012ApJ...758L...7R}.

On \textit{Episode 2} all technical, numerical and basic physical processes have been tested in the literature, and the fireshell model is now routinely applied to all GRBs, see e.g., GRB 101023 in \citet{2012A&A...538A..58P} and GRB 110709B in \citet{2013A&A...551A.133P}. 

The main aim of this article is dedicated to a deeper understanding of the physical and astrophysical process present in \textit{Episode 3}:

1) to evidence the universality properties of \textit{Episode 3}, observed in additional sources belonging to \textit{family 2}, as compared and contrasted to the very high variability of \textit{Episode 1} and \textit{Episode 2} components;

2) to present observations of GRB130427A leading to identify new physical regimes encountered in \textit{Episode 3} and their interpretation within the IGC paradigm;

3) to evidence the predictive power of the observations of \textit{Episode 3} and outline the underlying physical process leading to the characterization of the two above mentioned families of GRBs.

To start we will summarize in the next paragraph some qualifying new features generally observed in \textit{Episode 3} of selected GRBs of the \textit{Family 2} and proceed in the following paragraphs to the specific new informations acquired \textit{Episode 3} from GRB 130427A. We will then proceed to the general conclusions.

\begin{figure}
\centering
\includegraphics[width=\hsize]{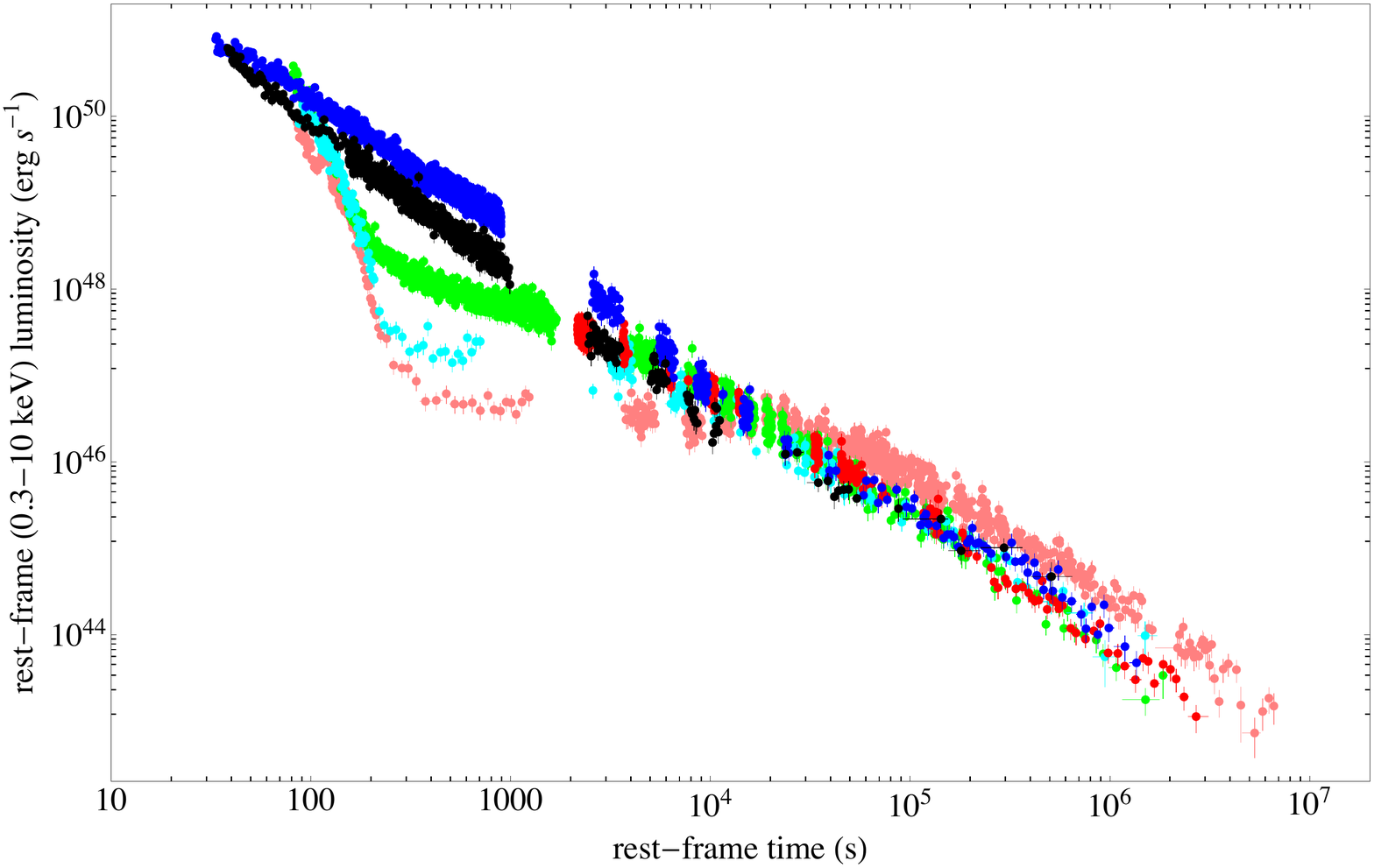}
\caption{The golden sample scaling law  \citep{Pisani:2013id}. X-ray luminosity light curves of the six GRBs with measured redshift in the $0.3−10$ keV rest-frame energy range: in pink GRB 060729, $z = 0.54$; black GRB 061007, $z = 1.261$; blue GRB 080319B, $z = 0.937$; green GRB 090618, $z = 0.54$, red GRB 091127, $z = 0.49$, and in cyan GRB 111228, $z = 0.713$.}
\label{pisani}
\end{figure}

\section{The qualifying features of \textit{Episode 3}}\label{sec:2}

As observations of additional sources fulfilling the IGC paradigm were performed, some precise qualifying features for characterizing \textit{Episode 3} have been found:

a) In some GRBs with known redshift belonging to this \textit{family 2} the late X-ray luminosities at times larger than $10^4$~s appeared to overlap, when duly scaled in the proper rest frame of the GRB source \citep{2012A&A...538A..58P}. This was soon confirmed for a sample of 6 GRBs, i.e. GRB 060729, GRB 061007, GRB 080319B, GRB090618, GRB 091127, GRB 111228, which we have called the  ``golden sample'' (\textbf{\emph{GS}}) \citep{Pisani:2013id}, see Figure \ref{pisani}. This unexpected result has led to adopt this universal luminosity versus time relation in the late X-ray emission of \textit{Episode 3} as a distance indicator. For some GRBs without a known cosmological redshift and exhibiting the general features of the four episodes, we imposed the overlapping of the late power law X-ray emission in their \textit{Episode 3} with the ones of the \textbf{\emph{GS}} and we have consequently inferred the value of the cosmological redshift of the source. This in turn has led to inferring the overall energetics of the source and to proceed to a consistent description of each episode following our theoretical model. This has been the case with GRB 101023, having inferred redshift $z=0.9$ and $E_{iso}=4.03 \times 10^{53}$ erg \citep{2012A&A...538A..58P}, and GRB 110709B, with inferred redshift $z=0.75$ and $E_{iso}=2.43\times10^{52}$ erg \citep{2013A&A...551A.133P}. 

\begin{figure}
\centering
\includegraphics[width=\hsize]{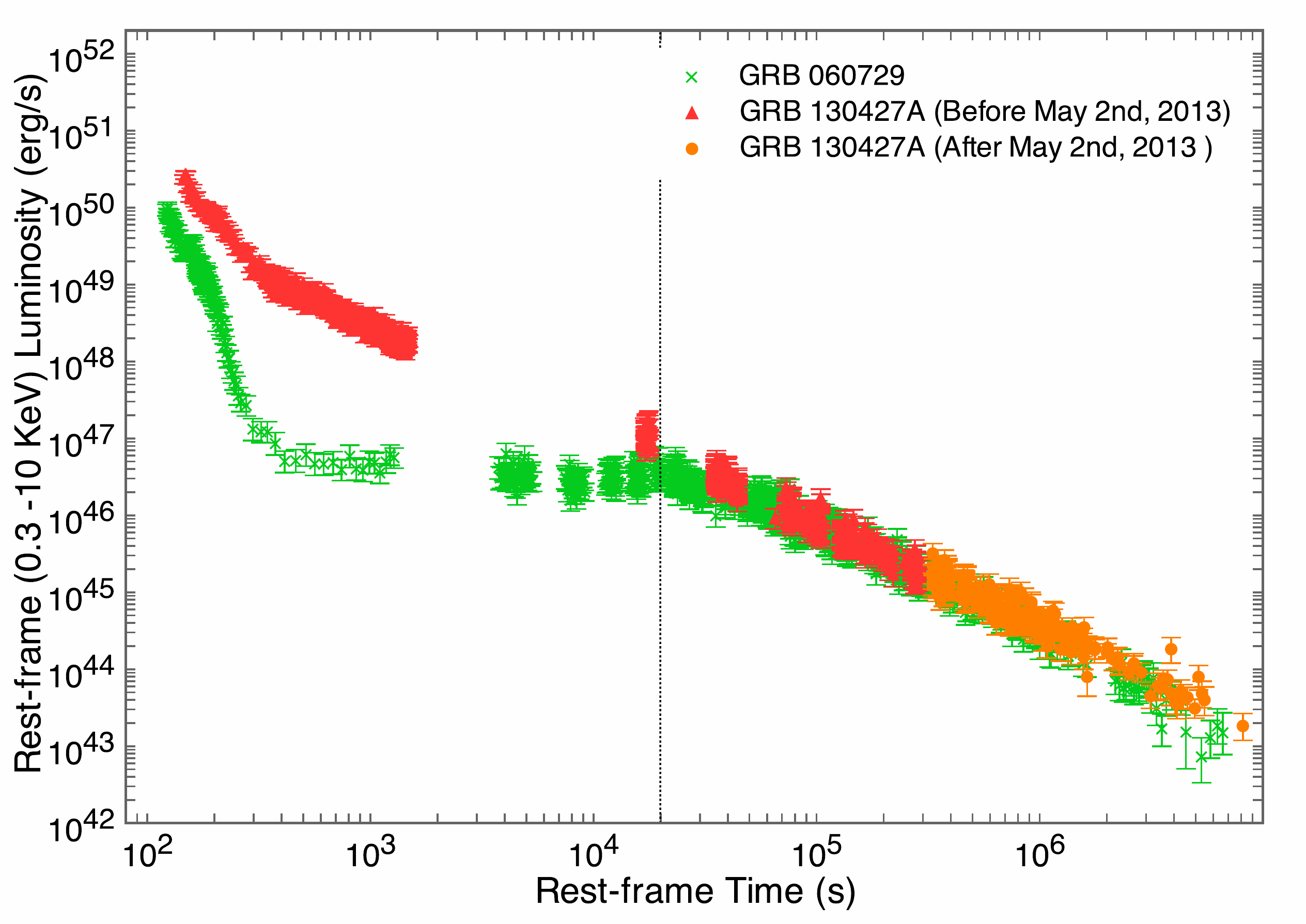}
\includegraphics[width=\hsize]{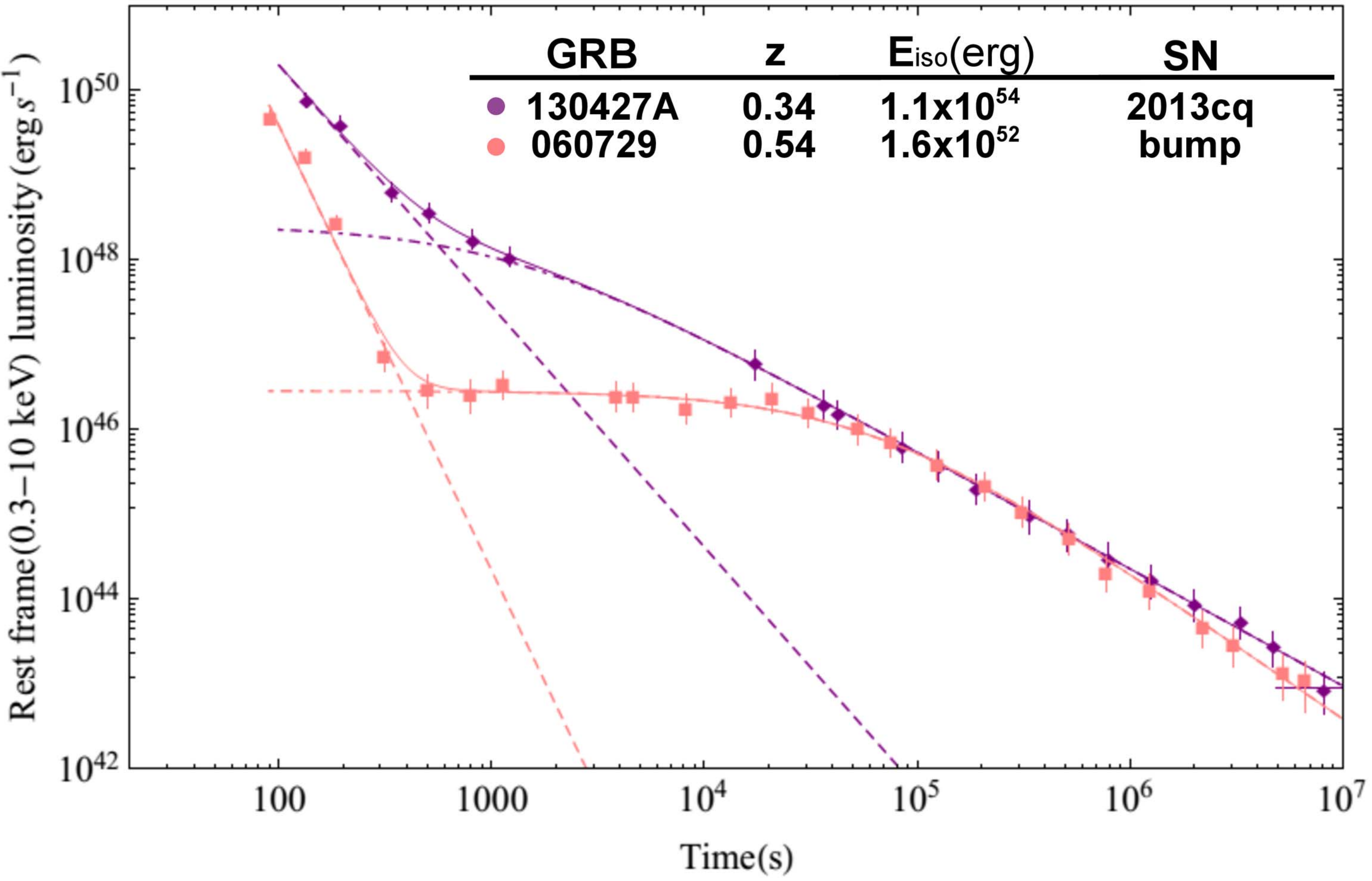}
\caption{\textbf{Top}: Overlapping of GRB 130427A and GRB 060729. Green cross is the light curve of GRB 060729. Red triangle and orange dots represent the light curve of GRB 130427A respectively before and after May 2, 2013. The vertical line marks the time of $2 \times 10^4$~s which is the lower limit for the domain of validity of the Pisani relation prior to GRB 130427A. \textbf{Bottom}: This figure shows GRB 060729 and 130427A have different magnitudes of the isotropic energy, but exhibit a common scaling law after $2\times10^4$~s. It also shows that the low isotropic energy GRB 060729 has a longer plateau, while the high isotropic energy GRB 130427A doesn't display an obvious plateau.}
\label{Nesting}
\end{figure}

The above analysis has initially addressed sources with $z<1$, where the associated SNe are observable. There is no reason to doubt that the IGC paradigm applies as well to sources for $z>1$. In this case clearly the SN is not observable with the current optical telescopes but the existence of all the above episodes, with the exception of \textit{Episode 4} related to the optical observation of the SN, in principle can be verified, if they are above the observational threshold, and the members of the \textbf{\emph{GS}} are correspondingly further increased. Indeed, significant results have been reached by observing the fulfillment of the above scaling laws in \textit{Episode 3} of GRB 090423, at $z=8.2$ \citep{2014arXiv1404.1840R}. The occurrence of this overlapping in the late X-ray emission observed by XRT has been considered as the necessary and sufficient condition to assert that a GRB fulfills the IGC paradigm.

b) The identification of a thermal emission occurring in the initial very steep decay of \textit{Swift}-XRT data of \textit{Episode 3} in GRB 090618 \citep{2014A&A...565L..10R}. We are currently examining other GRBs showing this feature, e.g., 060729, 061007, 061121 \citep{2011MNRAS.416.2078P,2012MNRAS.427.2950S,Friis:2013ef}. From these thermal emissions it is possible to infer the dimensions of the X-ray emitting regions as well as their Lorentz $\Gamma$ factors in this earliest part of \textit{Episode 3} \citep{2014A&A...565L..10R}. A typical mildly relativistic expansion regime with  $\Gamma \lesssim 2$ and characteristic radii $R \sim 10^{13}$ cm has been identified \citep{2014A&A...565L..10R}. These observational facts lead to a novel approach to the theoretical understanding of the X-ray emission process of \textit{Episode 3}, profoundly different from the ultra-relativistic one in the traditional jet afterglow collapsar paradigm model \citep{Piran:2005cs, Meszaros:2006gn}. We have concluded that this emission is not only mildly-relativistic, but also linked to a wide angle emission from the SN ejecta, in the absence of any sign of collimation \citep{2014A&A...565L..10R}.

c) From the direct comparison of the late X-ray emission of the \textbf{\emph{GS}} sources, we have recently identified the appearance of a ``nested structure'', which we illustrate in Figure \ref{Nesting}, comparing and contrasting the corresponding behavior of GRB 130427A with one the \textbf{\emph{GS}} GRB 060729 \citep{2014A&A...565L..10R}. The occurrence of these nested structures shows, among others, that in the case of the most intense sources, the common power law observed for the X-ray luminosities for time larger than $10^4$ s do extend to earlier times, see Figure  \ref{Nesting}. Indeed, for the most intense sources the common power law behavior is attained at an earlier time and at higher X-ray luminosities than the characteristic time scale indicated in \citep{Pisani:2013id}, see Figure \ref{pisani}. As we are going to show, in the present highly energetic GRB 130427A, this behavior starts at much earlier times around $400$~s.

Some of the above results were presented by one of the authors in the 2013 Texas Symposium on Relativistic Astrophysics \footnote{http://nsm.utdallas.edu/texas2013/proceedings/3/1/Ruffini.pdf}. There, referring to these sources originating in a tight binary system composed of a FeCO core at the onset of a SN event and a companion NS were named ``binary driven hypernovae'' \citep[BdHNe,][]{2014A&A...565L..10R}, in order to distinguish them from the traditional hypernovae (HN). 

The occurrence of the three features of \textit{Episode 3} listed above as obtained by our data analysis are becoming crucial to the theoretical understanding of the GRB-SN phenomena. They have never been envisaged to exist nor predicted in the traditional collapsar-fireball paradigm \citep{Nava:2013gf,vanEerten:2012hp,vanEerten:2010gw}. The IGC paradigm motivated an attentive data analysis of \textit{Episode 3} and the discovery of its universality has been a by-product.

\section{\textit{Episode 3} in the case of GRB 130427A}\label{sec:3}

We are going to show in this paper, in what follows, how GRB 130427A, associated with SN 2013cq and being the most luminous GRB ever observed in the past $40$ years, offers the longest multi-wavelength observations of \textit{Episode 3} so far. It confirms and extends all the above understanding and the corresponding scaling laws already observed in X-ray to lower and higher energies. It allows the exploration of the occurrence of similar constant power law emission in the high energy emission (GeV) and in the optical domain. We proceed with our data analysis of the ultra high GeV energy observations (\textit{Fermi}-LAT), those in soft and hard X-rays (\textit{Swift}-XRT and \textit{NuStar}, respectively) as well as of optical observations (\textit{Swift}-UVOT and ground based satellites). Our results are compared to and contrasted with the current ones in the literature. These observational facts set very specific limits: a) on the Lorentz $\Gamma$ factor of each component; b) on the corresponding mechanism of emission; c) on the clear independence of any prolongation of the GRB emission of \textit{Episode 2} to the emission process of \textit{Episode 3}.

The observation of the scaling law in the first $2 \times 10^4$~s alone has allowed us to verify the BdHN nature of this source which necessarily implies the presence of a SN. Consequently, we recall in Sec.~\ref{sec:3.1} that we made the successful prediction of the occurrence of a supernova which was observed in the optical band, as predicted on May 2, 2013.

In Sec.~\ref{sec:3.2}, we summarize our own data reduction of the \textit{Fermi} and \textit{Swift} satellites, we compare and contrast them with the ones in the current literature. 

In Sec.~\ref{sec:3.3}, we discuss the finding of a thermal component in the early part of X-ray emission of \textit{Episode 3}: this is crucial for identifying the existence of X-ray emission of a regime with low Lorentz factor and small radius, typical for supernova ejecta. 

In Sec.~\ref{sec:3.4} we compare and contrast the broad band (optical, X-ray, $\gamma$-rays all the way up to GeV) light curves and spectra of \textit{Episode 3} 

In Sec.~\ref{sec:3.5} we point out the crucial difference between the X-, $\gamma$-ray and GeV emission in Episode-3. 

In Sec.~\ref{sec:3.6} we proceed to a few general considerations on ongoing theoretical activities.

Secs.~\ref{sec:4} is the summary and conclusions.

\subsection{Identification and prediction}\label{sec:3.1}

With the appearance of GRB 130427A, we decided , as recalled above,  to explore the applicability of the IGC paradigm in the ``terra incognita'' of GRB energies up to $\sim 10^{54}$~erg. In fact, prior to GRB 130427A, the only known case of an equally energetic source, GRB 080319B, gave some evidence of an optical bump \citep{2009ApJ...691..723B, 2010ApJ...725..625T}, but in no way a detailed knowledge of the SN spectrum or type. We soon noticed in GRB 130427A the characteristic overlapping of the late X-ray decay in the cosmological rest frame of the source with that of GRB 060729, a member of the golden sample (in red in Fig.~\ref{Nesting}), and from the overlapping we deduced a redshift which was consistent with the observational value $z=0.34$ \citep{2013GCN..14455...1L}.

Therefore from the observations of the first $2 \times 10^4$ s, GRB 130427A has been confirmed to fulfill the IGC paradigm, and we conclude, solely on this ground, that a SN should necessarily be observed under these circumstances. We sent the GCN circular 14526\footnote{GCN 14526: The late X-ray observations of GRB 130427A by \textit{Swift}-XRT clearly evidence a pattern typical of a family of GRBs associated to supernova (SN) following the Induce Gravitational Collapse (IGC) paradigm (Rueda \& Ruffini 2012; Pisani et al. 2013). We assume that the luminosity of the possible SN associated to GRB 130427A would be the one of 1998bw, as found in the IGC sample described in Pisani et al. 2013. Assuming the intergalactic absorption in the I-band (which corresponds to the R-band rest-frame) and the intrinsic one, assuming a Milky Way type for the host galaxy, we obtain a magnitude expected for the peak of the SN of I = 22 - 23 occurring 13-15 days after the GRB trigger, namely between the 10th and the 12th of May 2013. Further optical and radio observations are encouraged.}  \citep{2013GCN..14526...1R} on May 2, 2013 predicting that the optical R-band of a SN will reach its peak magnitude in about 10 days in the cosmological rest-frame on the basis of the IGC paradigm, and we encouraged observations. Indeed, starting from May 13, 2013, the telescopes GTC, Skynet and HST discovered the signals from the type Ic supernova SN 2013cq \citep{2013GCN..14646...1D, 2013GCN..14662...1T, 2013GCN..14686...1L, 2013arXiv1307.5338L,Xu:2013ic}. We kept updating the X-ray \textit{Swift} data for weeks and we confirmed the complete overlapping of the late X-ray luminosities, in the respective cosmological rest frames, of GRB 130427A and GRB 060729 (in orange in Fig.~\ref{Nesting}). From these data it soon became clear that the power law behavior of the late time X-ray luminosity with index $\alpha \sim 1.3$ indicated in \citep{Pisani:2013id}, leading to the new concept of the ``nesting of the light curves'', started in this very energetic source already at $\sim10^2$~s following an initial phase of steeper decay \citep{2014A&A...565L..10R}.

Contrary to the traditional approach which generally considers a GRB to be composed of the prompt emission followed by the afterglow, both of which vary from source to source, the IGC paradigm for this \textit{family 2} has introduced the \textit{Episode 3} which shows regularities and standard late time light curves, largely independent of the GRB energy. It soon became clear that, with \textit{Episode 3}, we were starting to test the details of the physics and astrophysics of as yet unexplored regimes implied by the IGC paradigm: 1) a $\nu$-NS and the SN ejecta, originating in \textit{Episode 1}, 2) a newly formed BH originating in \textit{Episode 2}, and 3) the possible interaction among these components observable in the standard features of \textit{Episode 3}.

The joint observations of the \textit{Swift}, \textit{NuStar} and \textit{Fermi} satellites have offered the unprecedented possibility of clarifying these new regimes with the addition of crucial observations in the optical, X-ray and high energy radiation for \textit{Episode 3} of GRB 130427A, leading to equally unexpected results. The remainder of this article is dedicated to the understanding of \textit{Episode 3} of this remarkable event.

\subsection{Data Analysis of \textit{Episode 3} in GRB 130427A}\label{sec:3.2}

GRB 130427A was first observed by the \textit{Fermi}-GBM at 07:47:06.42 UT on April 27 2013 \citep{2013GCN..14473...1V}, which we set as the starting time $t_0$ throughout the entire analysis. After 51.1 s, the Burst Alert Telescope (BAT) onboard \textit{Swift} was triggered. The \textit{Swift} Ultra Violet Optical Telescope (UVOT) and the \textit{Swift} X-ray Telescope (XRT)  began observing at 181 s and 195 s after the GBM trigger respectively \citep{2013GCN..14448...1M}. Since this was an extremely bright burst, successively more telescopes pointed at the source: the Gemini North telescope at Hawaii \citep{2013GCN..14455...1L}, the Nordic Optical Telescope (NOT) \citep{2013GCN..14478...1X} and the VLT/X-shooter \citep{2013GCN..14491...1F} which confirmed the redshift $z = 0.34$.

GRB 130427A is one of the few GRBs with an observed adequate fluence in the optical, X-ray and GeV bands simultaneously for hundreds of seconds. In particular it remained continuously in the LAT field of view until 750 s after the trigger of \textit{Fermi}-GBM \citep{Ackermann:2013ih}, which gives us the best opportunity so far to compare the light curves and spectra in different energy bands, and to verify our IGC paradigm. We did the data reduction of \textit{Fermi} and \textit{Swift} satellites by the following methods.

\begin{figure}
\centering
\includegraphics[width=\hsize]{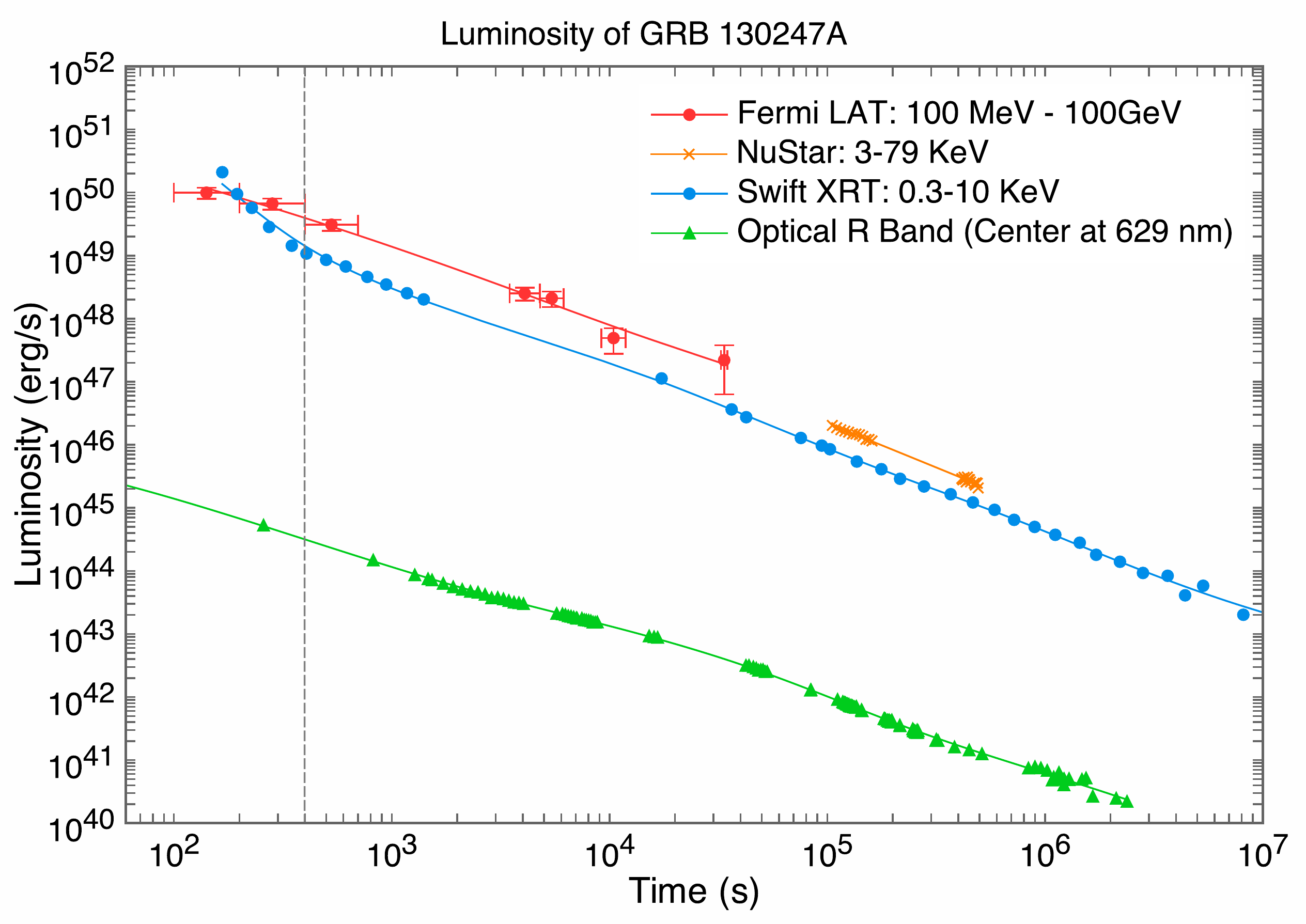}
\caption{The multi-wavelength light curve of GRB 130427A. The high energy (100~MeV--100~GeV) emission detected by \textit{Fermi}-LAT marked with red  and soft X-ray ($0.3$--$10$~keV) data from \textit{Swift}-XRT marked with blue are deduced from the original data. NuStar data ($3$~--~$79$~keV) marked with orange comes from \citep{Kouveliotou:2013jx}. The optical (R band, center at 629 nm) data marked with green comes from ground based satellites \citep{Perley:2013tf}. The error bars are too small with respect to the data points except for \textit{Fermi}-LAT data. The horizontal error bars of \textit{Fermi}-LAT represent the time bin in which the flux is calculated and vertical bars are statistical $1-\sigma$ errors on the flux (the systematic error of 10\% is ignored). The details in the first tens of seconds are ignored as we are interested in the behavior of the high energy light curve on a longer time scale. The vertical gray dashed line at ($\sim 400s$) indicates when the constant decaying slope starts. It is clear that all the energy bands have almost the same slope after 400~s in \textit{Episode 3}.}
\label{lc}
\end{figure}

\begin{figure}
\centering
\includegraphics[width=\hsize]{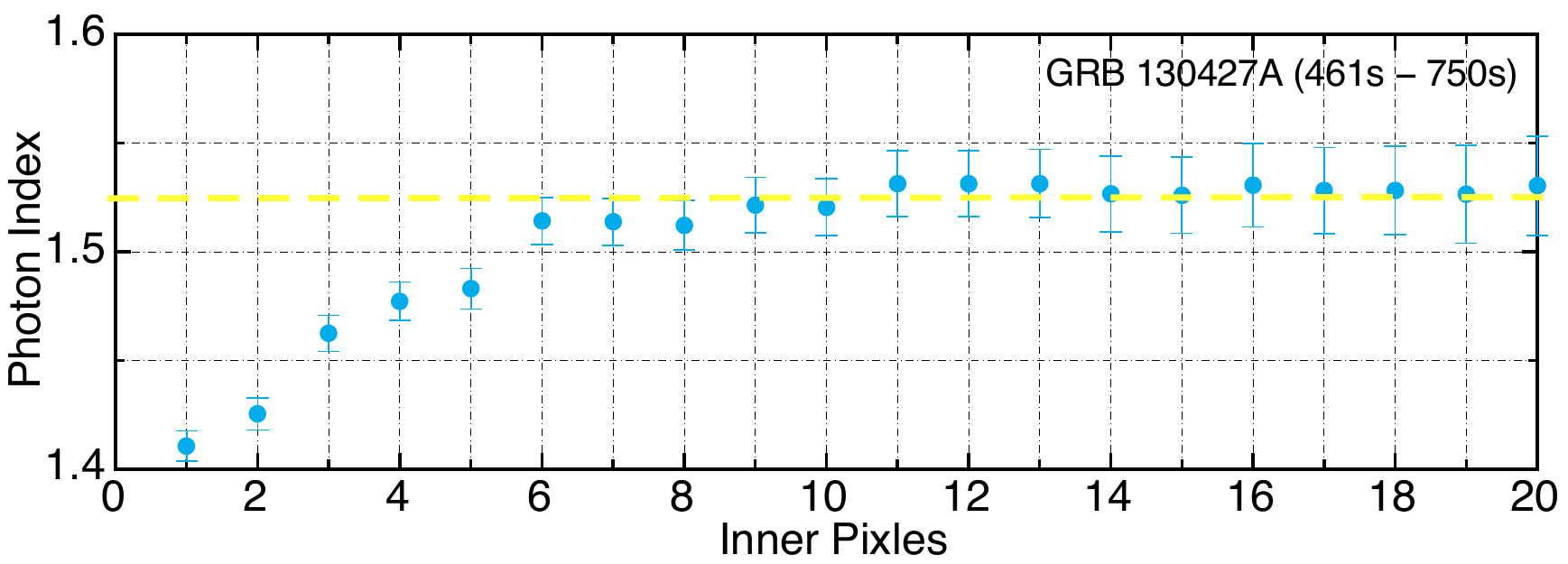}
\includegraphics[width=\hsize]{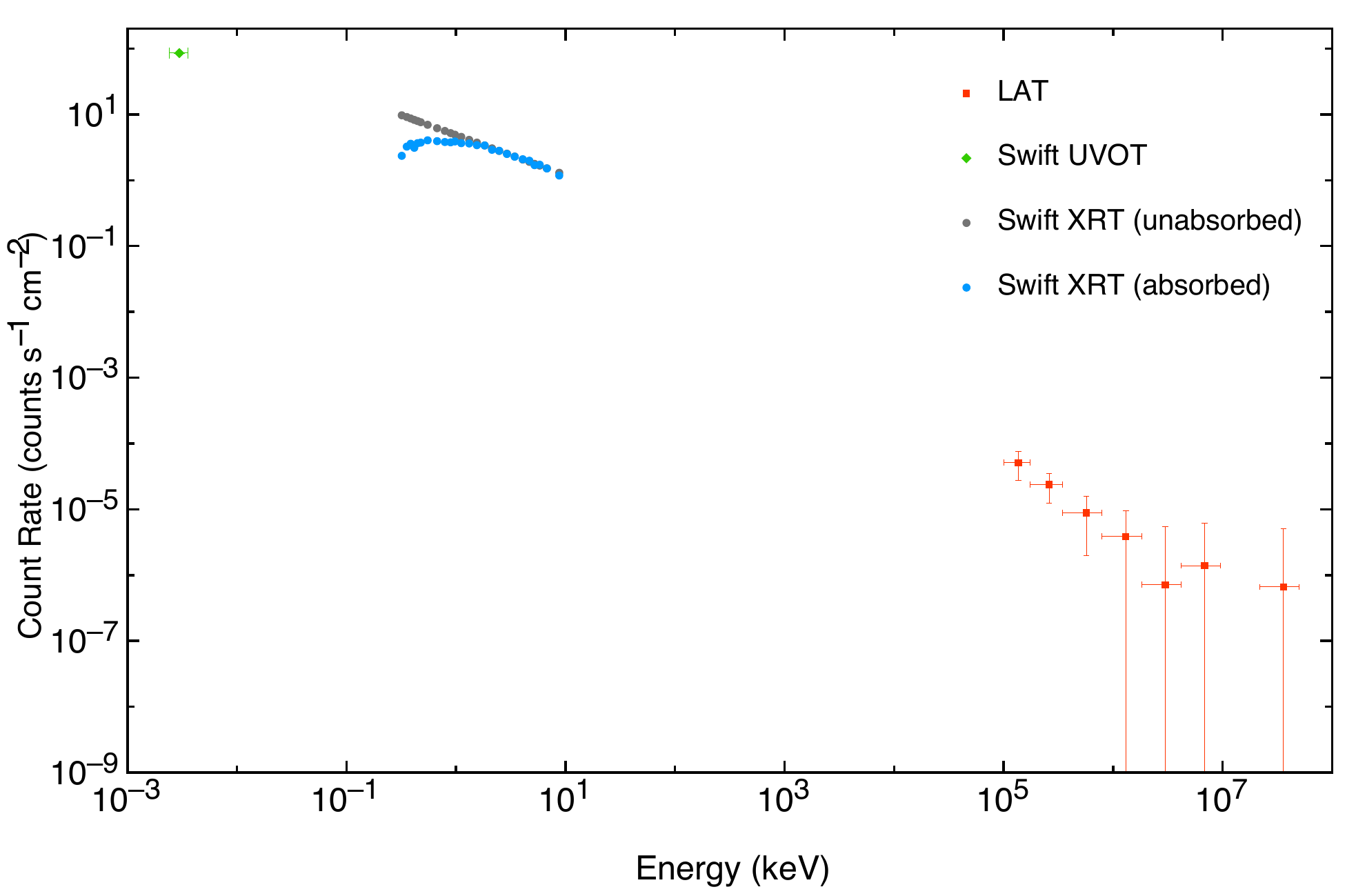}
\caption{Top: Data from the \textit{Swift}-XRT ($0.3$--$10$~keV) in the time range of 461--750~s for GRB 130427A. The data shows the photon index for different region selections after considering the pile up effect. After 6 inner pixels the photon index approaches an almost constant value of 1.52. Bottom: Spectra of GRB 130427A in the time range of 461--750~s. The green data points are from \textit{Swift}-UVOT \citep{Perley:2013tf}, the blue and gray points come from \textit{Swift}-XRT and red data correspond to \textit{Fermi}-LAT. The horizontal error bars are energy bins in which the flux is integrated and the vertical ones are $1-\sigma$ statistical errors on the count rate. The gray data points correspond to unabsorbed \textit{Swift}-XRT data while the blue ones are obtained with the assumption of absorption.}
\label{spec}
\end{figure}

\textit{Fermi}: Data were obtained from the Fermi Science Support Center\footnote{\noindent \url{http://Fermi.gsfc.nasa.gov}}, and were analyzed using an unbinned likelihood method with Fermi Science Tools v9r27p1\footnote{\noindent \url{http://Fermi.gsfc.nasa.gov/ssc/data/analysis/software/}}.
Event selections $P7SOURCE\_V6$ and $P7CLEAN\_V6$ were used, depending on which one gave more stable results. Recommended data cuts were used (e.g., $z_{max}=100$ degree). The background is composed of the galactic diffuse emission template and the isotropic emission template as well as about 60 point sources which are within the 15 degree radius of the GRB (however, their contribution was found to be negligible). The parameters for the background templates were held fixed during the fit. Luminosity light curve in Figure \ref{lc} corresponds to the energy range of 100 MeV to 100 GeV, circle radius of 15 degrees, with a power law spectra assumption. Since the data points up to the last two give a photon index of $\simeq 2.1$ with small errors, we set the photon index for the last two points to the value $2.1$ during the fitting procedure in order to obtain more stable results. The light curve can be obtained with great temporal detail before $750$ s. However, since we are interested in the general behavior of \textit{Episode 3}, for simplicity we neglected such a fine temporal structure and we rebinned the light curve. Therefore there are only 3 data points up to $750$ s. The spectrum is plotted in Figure \ref{spec}.

\textit{Swift}: XRT data were retrieved from UKSSDC \footnote{\noindent \url{http://www.Swift.ac.uk}} and were analyzed by the standard Swift analysis software included in the NASA's Heasoft 6.14 with relevant calibration files\footnote{\noindent \url{http://heasarc.gsfc.nasa.gov/lheasoft/}}. In the first 750~s only Windows Timing (WT) data exists and the average count rate exceeds 300~counts/s: the highest count rate even reaches up to 1000~counts/s, far beyond the value of 150~counts/s which is suggested for the WT mode as a threshold of considering pile-up effects \citep{Evans:2007iz}. Pile-up effects cause the detector to misrecognize two or more low energy photons as a single high energy photon, which softens the spectrum. We adopted the method proposed by \citet{Romano:2006kt}, fitting dozens of spectra from different inner sizes of box annulus selections in order to determine the extent of the distorted region. Taking the time interval 461~s to 750~s as an example, the deviation comes from where the inner size is smaller than 6 pixels, shown in Figure \ref{spec}. Then we applied the standard XRT data analyzing process \citep{Evans:2007iz,Evans:2009kx} to obtain the spectrum, plotted in Figure \ref{spec}. For the luminosity light curve, we split XRT observations in the nominal $0.3$--$10$~keV energy range to several slices with a fixed count number, and we followed the standard procedure \citep{Evans:2007iz,Evans:2009kx} and considered the pile-up correction. The XRT light curves of different bands are shown in Figure \ref{lc}.

\subsection{The X-ray qualification of GRB 130427A as a BdHN}\label{sec:3.3}

Here we first focus on the extended X-ray emission of \textit{Episode 3} which, as we have shown above, gives the qualifying features for the identification of GRB 130427A as a BdHN. 
We first proceed to identify the power law component of the light curve after the steep decay and the end of the plateau. This power law component, in the present case of this most energetic source GRB 130427A, has a power law index $\alpha = -(1.31\pm 0.01)$ and it extends all the way from $400$~s to $\sim 10^7$~s without jet breaks. These results are consistent with some previous papers \citep[see e.g. in][]{Perley:2013tf,2013ApJ...776..119L} which find no jet break, but differs from \citep{Maselli:2013hc} in which a break of the later time light curve is claimed. 

\begin{figure}
\centering
\includegraphics[width=\hsize]{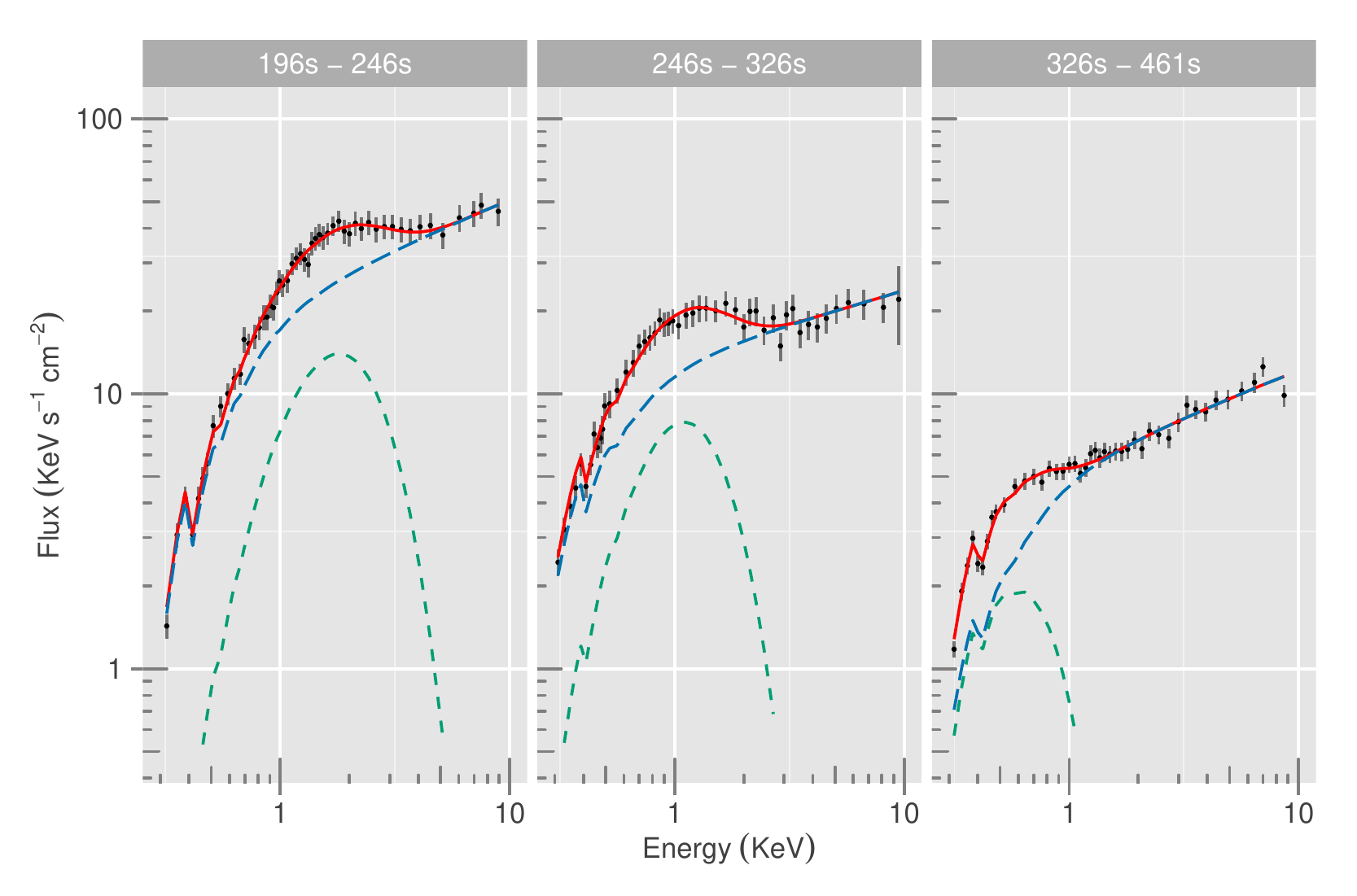}
\caption{Spectral fitting of three time intervals (196s - 246s, 246s - 326s, 326s - 461s) in \textit{Episode 3}, data come from \textit{Swift}-XRT (0.3 KeV - 10 KeV, without pile-up area). Black points are the deduced data, green dashed line presents the thermal component, blue long-dashed line is the power law component, and red line shows the combination of these two components. Clearly the flux of thermal component drops and the temperature decreases along the time.}
\label{pobb}
\end{figure}

We turn now to an additional crucial point: to confirm that the X ray  emission of \textit{Episode 3} belongs to the SN ejecta and not to the GRB. To do this it is crucial, as already done for other sources \citep{2014A&A...565L..10R}, to determine the presence of a thermal component in the early time of \textit{Episode 3} and infer its temperature and the size of its emitter. Indeed, by analyzing the XRT data, we find that adding a blackbody component efficiently improves the fit with respect to a single power law from 196~s to 461~s. The corresponding blackbody temperature decreases in that time duration from $0.5$~keV to $0.1$~keV, in the observed frame. Figure \ref{pobb} shows the evolution of the power law plus blackbody spectra in three time intervals, clearly the flux of thermal component drops along the time, as well as the temperature corresponding to the peak flux energy decreases.
\citet{Kouveliotou:2013jx} find that a single power law is enough to fit the \textit{NuStar} data in the \textit{NuStar} epochs, the reason could be that the thermal component has faded away or exceeded the observational capacity of the \textit{Swift} satellite in the \textit{NuStar} epochs, which start later than $10^5$~s. 

By assuming that the blackbody radiation is isotropic in the rest frame, the emitter radius along the light of sight increases from $\sim 0.7\times10^{13}$~cm at 196~s to $\sim 2.8\times10^{13}$~cm at 461~s in the observed frame, orders of magnitude smaller than the emission radius of the GRB, which is larger than $10^{15}$~cm in the traditional GRB collapsar afterglow model. The size of $10^{13}$~cm at hundreds of seconds is consistent with the observation of  supernova ejecta. After considering the cosmological and the relativistic corrections, $t_a^d \simeq t (1+z)/{2 \Gamma ^2}$, where $t$ and $t_a^d$ are the time in the laboratory and observed frame respectively, and $\Gamma$ is the Lorentz factor of the emitter, we get an expansion speed of $\sim 0.8c$, corresponding to Lorentz factor $\Gamma = 1.67$. These results contradict  the considerations inferred in \citep{Maselli:2013hc} $\Gamma \sim 500$, which invoke a value of the Lorentz factor in the traditional collapsar afterglow model \citep[see e.g.][]{Meszaros:2006gn}. Again in the prototypical GRB 090618, the Lorentz factors ($1.5 \leq \Gamma \leq 2.19$) and emission radii ($\sim 10^{13}$~cm) are very similar to the ones of GRB 130427A presented in \citet{2014A&A...565L..10R}. It is interesting that such a thermal component has been also found in the early parts of \textit{Episode 3} of GRB 060729 (adopted in Fig. \ref{Nesting}) and many other SN associated GRBs \citep[see][]{2014A&A...565L..10R, 2007ApJ...662..443G, 2012MNRAS.427.2950S}.

\subsection{Discussion of multi-wavelength observations in \textit{Episode 3}}\label{sec:3.4}

\begin{figure}
\centering
\includegraphics[width=\hsize]{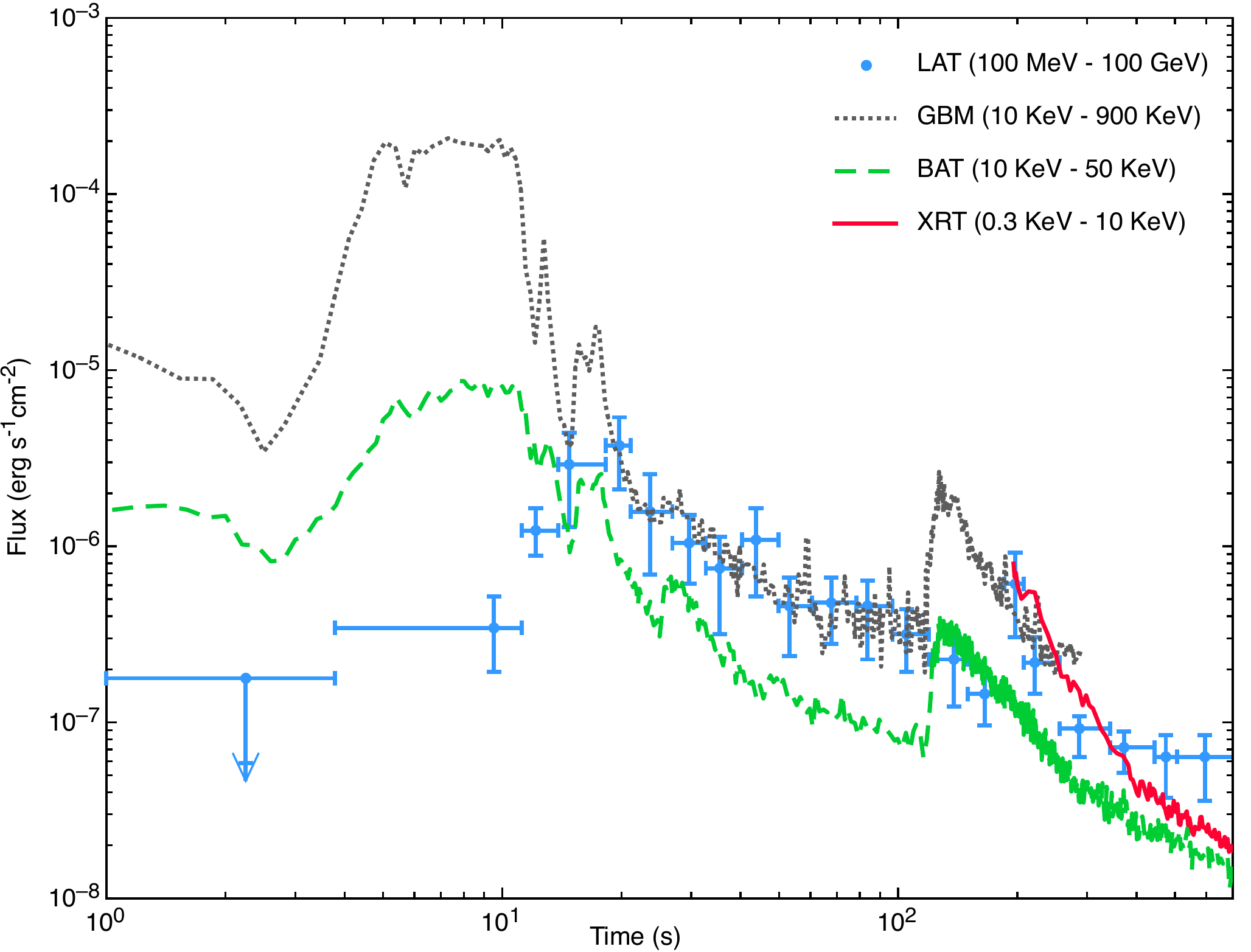}
\caption{Flux of first 700~s. Blue points are the \textit{Fermi}-LAT high energy emission from $100$~MeV till $100$~GeV \citep{Ackermann:2013ih}, grey dotted line represents the \textit{Fermi}-GBM, from $10$~keV to $900$~keV,  green dashed line represents the photons detected by \textit{Swift}~BAT from $10$~keV to $50$~keV, and red solid line is the soft X-ray \textit{Swift}-XRT detection, in the range of $0.3$~KeV to $10$~KeV. From this figure, clearly the \textit{Fermi}-LAT emission reaches highest fluence at about 20~s while the gamma-ray detected by \textit{Fermi}-GBM releases most of the energy within the first 10~s.}
\label{gev}
\end{figure}

Now we turn to the most unexpected feature in the analysis of the optical, X-ray, $\gamma$-ray and very high energy emission in \textit{Episode 3} of GRB 130427A.
The optical emission was observed by \textit{Swift}-UVOT and many ground-based telescopes (R band as an example for the optical observation).  The soft X-ray radiation was observed by \textit{Swift}-XRT
($0.3$--$10$~keV).  Similarly the hard X-ray radiation was observed by \textit{Swift}-BAT
($15$--$150$~keV) and by \textit{NuStar} ($3$--$79$~keV).  The gamma ray radiation was observed by \textit{Fermi}-GBM ($8$~KeV --$40$~MeV), and the high energy radiation by \textit{Fermi}-LAT ($100$~MeV~--~$100$~GeV).
The main result is that strong analogies are found in the late emission at all wavelengths in \textit{Episode 3}: after 400s, these luminosities show a common power law behavior with the same constant index as in the X-ray (and clearly with different normalizations), by fitting multi-wavelength light curves together we have a power law index $\alpha = -(1.3\pm 0.1)$.

Turning now to the spectrum, integrated between $461$~s and $750$~s, the energy range covers $10$ orders of magnitude, and the best fit is a broken power law (see Fig.~ \ref{spec}). In addition to the traditional requirements for the optical supernova emission in \textit{Episode 4}, the much more energetically demanding requirement for the general multi-wavelength emission of \textit{Episode 3} has to be addressed.

\subsection{The onset of \textit{Episode 3}}\label{sec:3.5}

In the previous sections we have emphasized the clear evidence of GeV emission and its analogy in the late power law luminosities as functions of the arrival time for the X-ray, optical and GeV emissions.
Equally important in this section is to emphasize some differences between the X-, $\gamma$-ray, and the high energy GeV emission, especially with respect to the onset of \textit{Episode 3} at the end of prompt emission in \textit{Episode 2} (see Fig.~\ref{gev}). We observe:

1) The $\gamma$-ray light curves, observed by \textit{Fermi}-GBM and hard X-ray observed by \textit{Swift}-BAT, have similar shapes. They reach the highest luminosity between 4~s to 10~s during the prompt emission phase of \textit{Episode 2}.

2) The high energy ($>100$ MeV) GeV emission gradually rises up, just after the gamma and X-ray prompt emissions drop down at the end of Episode-2: the high energy GeV emission raises to its peak luminosity at about 20~s.The turn on of the GeV emission coincides, therefore, with the onset of our Episode~3. These considerations have been recently confirmed and extended by the earliest high energy observations in GRB 090510 (Ruffini et al. 2014, submitted to ApJL).

3) At about 100~s, the \textit{Swift}-XRT starts to observe the soft X-ray and a sharp spike appears in the hard X-ray and gamma ray bands (see Fig.~\ref{gev}). Only at this point the \textit{Swift}-XRT started to observe soft X-ray. We are currently addressing the occurrence of the spike to the thermal emission observed to follow in the sharp decay of the X-ray luminosity prior to the plateau and the above mentioned common power law decay \citep{Ruffini2014c}.

\begin{figure}
\centering
\includegraphics[width=\hsize]{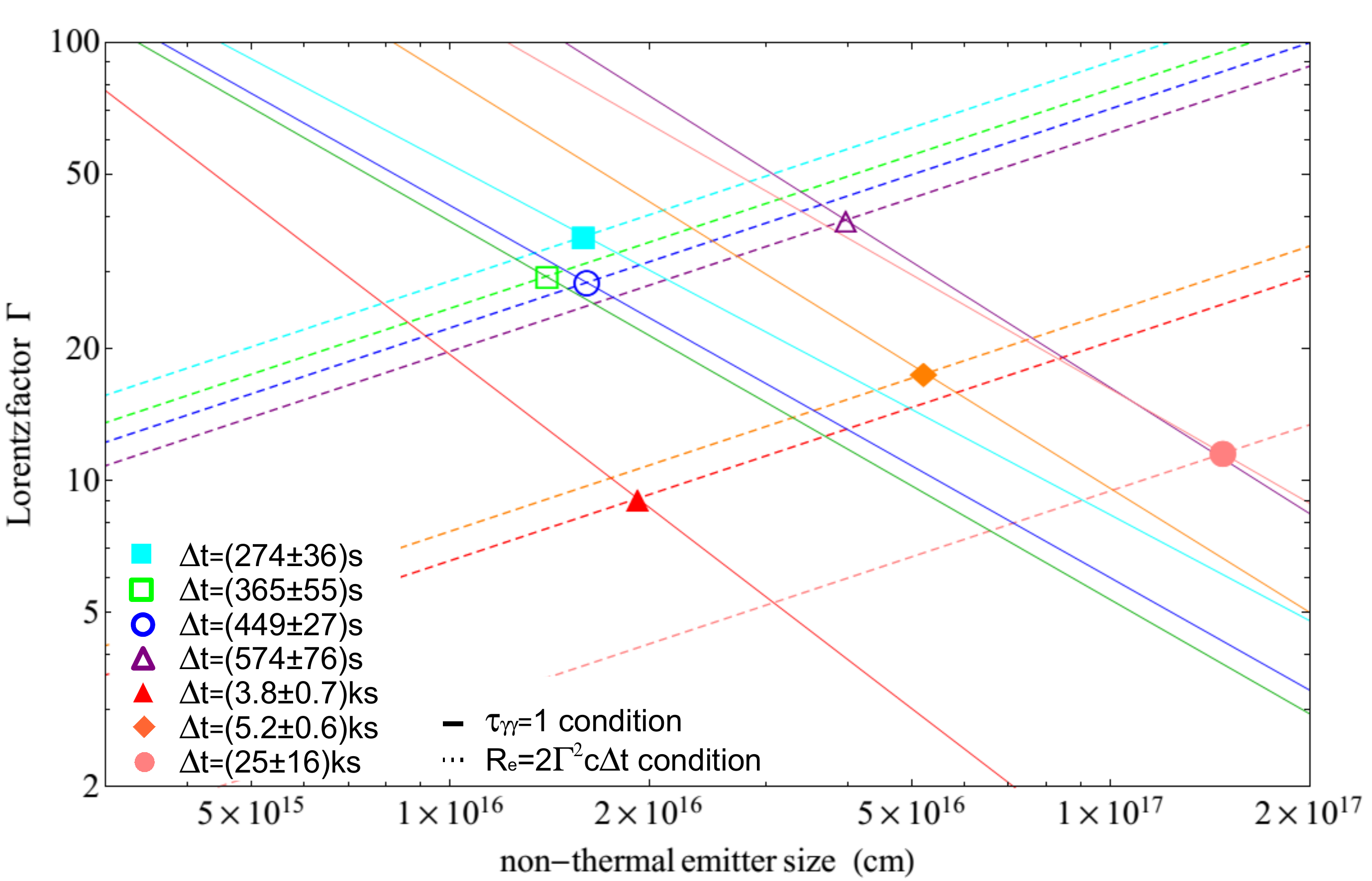}
\caption{Constraints on the Lorentz factors and on the size of the GeV emitting region at the transparency point. Solid curves represent the curves defined by varying the emitting region size from the $\tau_{\gamma\gamma}=1$ condition; dot-dashed curves represent the radius of the emitter obtained from causality in the ultra-relativistic regime, i.e., $R_{em}=2\Gamma^2c\Delta t$. Filled circles correspond to the solutions of both the limits. The different colors refer to the time intervals from $\sim273$ s to $24887$ s, in the order: cyan, green, blue, purple, red, orange, and pink.}
\label{gammagev}
\end{figure}

The detailed analysis of the prolonged emission observed by \textit{Fermi}-LAT in GeV enables us to set specific limits on the Lorentz factor of this high energy emission. 
We have analyzed the GeV emission from $\sim300$ s to $2.5 \times 10^4$ s, dividing this time interval into seven sub-intervals and  in each of them collecting the corresponding maximum photon energy and photon index of the spectral energy distribution, as shown in \citet[][Fig.~2]{Ackermann:2013ih}. We have focused our attention on the estimate on the Lorentz factor for this high energy component from the usual optical depth formula for pair creation $\tau_{\gamma\gamma}$ \citep[see, e.g.,][]{Lithwick:2001vg,Gupta:2008cc}. We have computed for different values of radii of the emitter, the corresponding Lorentz factors at the transparency condition, i.e., $\tau_{\gamma\gamma}=1$, see the solid curves in Figure \ref{gammagev}. The constraints on the size of the  emitting regions come from causality in the ultra-relativistic regime, i.e., $R_{em}=2\Gamma^2c\Delta t$, where $\Delta t$ corresponds to the duration of the time intervals under consideration (see the dot-dashed curves in Figure \ref{gammagev}). The values of the Lorentz factor ranges between $\sim10$ and $\sim40$ and correspondingly, the radii of the emitting region at the transparency point are located between $\sim10^{16}$ cm and $\sim2\times10^{17}$ cm (see the filled circles in Figure \ref{gammagev}).

\subsection{General considerations on recent theoretical progress on BdHD}\label{sec:3.6}

The concurrence of the above well-defined scaling laws and power law of the observed luminosities both in the X-ray and/or in the optical domains in the \textit{Episode 3} of GRB 130427A have been considered  arguments in favor for looking to the $r$-process and to heavy nuclei radioactive decay as the energy sources \citep{2014A&A...565L..10R},\citep[see the pioneering work of][]{1998ApJ...507L..59L}. The extended interaction of the $\nu$-NS and its binary NS companion in the SN ejecta provides environment for $r$-processes to create the needed neutron rich very heavy elements to attribute some of the electromagnetic energy in \textit{Episode 3} to nuclear decay, $\approx 10^{52}$~erg . 
 Alternatively, we are considering emission originating from type-I and type-II Fermi acceleration mechanisms, introduced by Fermi precisely to explain the radiation process in the SN remnants \citep{Fermi:1949jk}. Also these processes can lead to a power law spectrum \citep{2004vhec.book.....A}, similar to the one presented in this article and in our recent letter \citep{2014A&A...565L..10R}. The GRB emission  of \textit{Episode 2} interacting with the Supernovae ejecta could represent that energy injection long sought by Fermi for the onset of his acceleration mechanism \citep{Fermi:1949jk}. 
 
Both of the above processes can indeed operate as energy sources for the mildly relativistic X-ray component and the relativistic GeV emission of \textit{Episode 3}.

We are currently examining additional BdHN sources and giving particular attention to understanding the highest GeV energy emission, which is unexpected in the traditional $r$-process. The inferred $\Gamma$ Lorentz factor for the GeV emission point to the possibility  of a direct role of the two remaining components in the IGC paradigm: the newly born neutron star ($\nu$NS) and the just born Black Hole (see Fig.~\ref{matrix}) There is also the distinct possibility that these two systems have themselves become members of a newly born binary system\footnote{Presentation of R.Ruffini in Yerevan: http://www.icranet.org/  images/ \\ stories/Meetings/meetingArmenia2014/talks/ruffini-1.pdf} \citep{2012ApJ...758L...7R}.

\section{Conclusion}\label{sec:4}

We have recalled that GRB 130427A is one of the most energetic GRBs ever observed ($E_{iso} \simeq 10^{54}$~ergs), with the largest $\gamma$-ray fluence and the longest lasting simultaneous optical, X-ray, $\gamma$-ray and GeV observations of the past 40 years.
For this reason we have performed our own data analysis of the \textit{Swift} and \textit{Fermi} satellites (see Secs.~\ref{sec:3}) in order to probe the BdHN nature of this source (see Sec.~\ref{sec:3.3}) and infer new perspectives for the IGC paradigm and the physical and astrophysical understanding of GRB.

We summarize the main results by showing how the analysis of GRB 130427A should be inserted in a wider context with three different areas: a)the formulation and the observational consequences of the IGC paradigm; b) the comprehension induced by the multi-wavelength observations of GRB130427A; c) the BDHN versus HN properties. Relevance of BDHN in establishing a new alternative distance indicator in astrophysics. Each one of these topic is going to be summarized in 3 bullets.

With reference to the formulation and observational consequences of the IGC paradigm:

1a) The IGC paradigm introduces in Astrophysics a new experience which has been already successfully applied in particle physics: to understand that a system traditionally considered elementary is actually a composed system and that new components in the system can appear at the effect of collisions or decay. Well known physical examples are represented by the introduction of the quark \cite{2012PhLB..716....1A}, or the creation of new particles in a decay or collision of elementary particle systems: the Fermi theory of beta decay or the mesons production in en electron positron collision in storage rings are classical examples. These facts are today routinely accepted in particle physics although Fermi had to spend efforts to explain them at the time \citep{Fermi:1934}. In astrophysics this situation is new: to see that a process until recently considered elementary, as the GRB, does indeed contains four different astrophysical systems and that two of them, the FeCO core undergoing SN and the companion NS binary, interacting give origin to two different new system a $\nu$NS and a BH and especially that the entire process occur in less the 200 s is a totally new condition. For its understanding a new approach technically and conceptually is needed. The new style of research is more similar to the one adopted in particle physics then the one in classical astronomy, see Figure \ref{matrix}.

2a) Possibly the most profound novelty in this approach, for the understanding of GRBs, has been the introduction of the four episodes shortly summarized in  Sec.\ref{sec:1}. The traditional GRB description corresponds to \textit{Episode 2}. \textit{Episode 1} corresponds to the dynamical accretion of the SN ejecta onto the companion NS. We are now considering an enormous rate of accretion of $10^{31}$~g~s$^{-1}$, $10^{15}$ times larger then the one usually considered in binary X-ray source in systems like Centaurus X-3 or Cygnus X-1 \citep[see e.g., in][]{Giacconi1975}. This process has opened a new field of research by presenting the first realization of the hypercritical accretion introduced by Bondi-Hoyle-Littleton as well the testing ground of the neutrinos emission pioneered in the \citet{1972SvA....16..209Z} and \citet{1975PhRvD..12.2959R}, see Sec. \ref{sec:1}. The pure analytic simplified solutions in \citet{2012ApJ...758L...7R} have been now supported by direct numerical simulation in \citet[][and Fig.~1 therein]{2014arXiv1409.1473F}.

3a) The main revolution of the IGC paradigm for GRBs comes from the discovery of the universal laws discovered in \textit{Episode 3} which compare and contrast to the explosive, irregular phases, varying from source to source in all observed GRBs, in their \textit{Episode 1} and \textit{Episode 2}.The universality of \textit{Episode 3} as well as the precise power laws and scaling laws discovered changes the field of GRB analysis making it one of time resolved, high precision,  reproducible measurements. Additional unexplored physical phenomena occurs in \textit{Episode 3}, adding to the new ultra-relativistic regimes already observed in the \textit{Episode 2} in previous years\footnote{Presentation of R.Ruffini in 13th Marcel Grossmann Meeting: http://www.icra.it/mg/mg13}, see Figure \ref{pisani} and Figure \ref{Nesting} as well as Figure \ref{lc}.

With reference to the comprehension induced by the multi-wavelength observations of GRB130427:

1b) Following the work on the \textbf{\emph{GS}} \citep{Pisani:2013id} and the more recent work on the nested structures \citep{2014A&A...565L..10R}, we have first verified that the soft X-ray emission of GRB 130427A follows for time $t \simeq 10^4$~s the power-law decay described in \citet{Pisani:2013id}. Surprisingly in this most energetic GRB unveils such power-law behavior already exists at the early time as t $\sim 100$~s (details in \citet{2014A&A...565L..10R}). From the X-ray thermal component observed at the beginning of \textit{Episode 3} following a spiked emission at $\sim 100$~s, a small Lorentz factor of the emitter is inferred ($\Gamma < 2$): this X-ray emission appear to originate in a mildly relativistic regime with a velocity  $v \sim 0.8c$, in addition it does not appear to have substantial beaming and appears to be relatively symmetric and with no jet break, see Figure \ref{lc}, and \citet[][Figure 2]{2014A&A...565L..10R}.  

2b) We have proceeded to make a multi-wavelength analysis of \textit{Episode 3} where we have compared and contrasted optical data from \textit{Swift}-UVOT and ground based telescopes, X-ray data from \textit{Swift}-XRT, $\gamma$-ray data from \textit{Fermi}-GBM, as well as very high energy data in the GeV from \textit{Fermi}-LAT. The high energy emission appears to be detectable at the end of the prompt radiation phase in \textit{Episode 2}, when the fluence of X-ray and $\gamma$-ray of the prompt exponentially decrease and becomes transparent for the very high energy photons in the \textit{Fermi}-LAT regime. From the transparency condition of the GeV emission a Lorentz Gamma factor of $10-40$ is deduced. In principle this radiation, although no brake in its power law is observed, could be in principle beamed, see Figure \ref{gammagev}. 

3b) Although the light curves of X-ray and GeV emission appear to be very similar, sharing similar power-law decay index, their Lorentz $\Gamma$ factors appear to be very different, and their physical origin are necessarily different. Within the IGC model the X-ray and the high energy can originate from the interaction of some of the physical components (e.g., neutron star and black hole) newly created in the C-matrix: the interaction of the GRBs with the SN ejecta \citep{Ruffini2014c} may well generate the X-ray emission and the associated thermal component. The high energy should be related to the novel three components, the BH, the $\nu$NS and the SN ejecta. From the dynamics it is likely that the $\nu$NS and the BH form a binary system, see e.g., \citet{2012ApJ...758L...7R} and the presentation by one of the authors\footnote{Presentation of R.Ruffini in Yerevan: http://www.icranet.org/images/ \\ stories/Meetings/meetingArmenia2014/talks/ruffini-1.pdf}.

With reference to the the BDHN versus HN properties:

1c) The verification of the BdHN paradigm in GRB 130427A has confirmed that for sources with isotropic energy approximately $10^{54}$~erg, the common power law behavior is attained at earlier times, i.e., $\sim 10^3$~s, and higher X-ray luminosities than the characteristic time scale indicated in \citep{Pisani:2013id}, see Figure ~\ref{pisani}. From the observation of the constant-index power law behavior in the first $2 \times 10^4$~s of the X-ray luminosity light curve, overlapping with the known BdHNe, it is possible to have an estimate of: 1) the redshift of the source, 2) the isotropic energy of the GRB and 3) the fulfillment of the necessary and sufficient condition for predicting the occurrence of the SN after $\sim10$~days in the rest frame of the source, see e.g., GCN 14526. This procedure has been successfully applied to GRB 140512A (Ruffini et al. in preparation).

2c) The overlapping with the \textbf{\emph{GS}} members of the late X-ray emission observed by swift XRT, referred to the rest frame of the source, introduces a method to establish and independent distance estimator of the GRBs. Although this method has been amply applied (e.g. GRBs 060729, 061007, 080319, 090618, 091127, 111228A), we also declare that there are some clear outlier to this phenomenon: GRB 060614 \citep{RuffiniGCN1}, 131202A \citep{RuffiniGCN2} and 140206A \citep{RuffiniGCN3}. These are all cases of great interest and the solution of this contradictions my reveal to be of particular astrophysical significance. Particularly interesting is the case of GRB 060614 since the cosmological redshift has not been directly measured and there can be a misidentification of the host galaxy \citep{2006ApJ...651L..85C}.

3c) As first pointed out in \citet{2012ApJ...758L...7R} and \citet{2014A&A...565L..10R}, further evidenced in \citet{2014arXiv1409.1473F}, the crucial factor which may explain the difference between the \textit{family 1} and \textit{family 2} of GRBs is the initial distance between the FeCO core and its binary NS companion. The accretion from the SN ejecta onto the companion NS and the consequent emission process decrease by increasing this distance: consequently is hampered the possibility for the binary companion NS to reach its critical mass \citep[see Fig.~3 and Fig.~4 in][and the discuss therein]{2012A&A...548L...5I}. Unlike \textit{family 2}, in \textit{family 1} no BH is formed, no GRB is emitted, and no \textit{Episode 2} nor \textit{Episode 3} exist, only a softer and less energetic radiation from the accretion onto the neutron star will be observed in these sources. The problem of explaining the coincidence between the GRB and supernova in the case of the \textit{family 1} is just a tautology: no GRB in this family exist but only a hypernova \citet{2014A&A...565L..10R}.

This article addresses recent results on the IGC paradigm applied to long GRBs. The IGC paradigm and the merging of binary neutron stars has been also considered for short GRBs \citep[see e.g.,][]{Muccino:2013cq,Muccino:2013et,Muccino:2014ex,2014A&A...565L..10R} and is now being further developed.

\begin{acknowledgements}

We acknowledge the use of public data from the \textit{Swift} and \textit{Fermi} data achieve. We thank Robert Jantzen for his careful reading and valuable advise. We also thank the editor and the referee for their constant attention and beneficial suggestions which largely improve this article. ME, MK and GBP are supported by the Erasmus Mundus Joint Doctorate Program by Grant Numbers 2012-1710, 2013-1471 and 2011-1640, respectively, from the EACEA of the European Commission.

\end{acknowledgements}


\begin{thebibliography}{}
\expandafter\ifx\csname natexlab\endcsname\relax\def\natexlab#1{#1}\fi

\bibitem[{Aad {et~al.}(2012)Aad, Abajyan, Abbott, Abdallah, Abdel~Khalek,
  Abdelalim, Abdinov, Aben, Abi, Abolins, AbouZeid, Abramowicz, Abreu, Acharya,
  Adamczyk, Adams, Addy, Adelman, Adomeit, Adragna, Adye, Aefsky,
  Aguilar-Saavedra, Agustoni, Aharrouche, Ahlen, Ahles, Ahmad, Ahsan, Aielli,
  Akdogan, {\AA}kesson, Akimoto, Akimov, Alam, Alam, Albert, Albrand, Aleksa,
  Aleksandrov, Alessandria, Alexa, Alexander, Alexandre, Alexopoulos, Alhroob,
  Aliev, Alimonti, Alison, Allbrooke, Allport, Allwood-Spiers, Almond, Aloisio,
  Alon, Alonso, Alonso, Altheimer, {\'A}lvarez~Gonz{\'a}lez, Alviggi, Amako,
  Amelung, Ammosov, Amor Dos~Santos, Amorim, Amram, Anastopoulos, Ancu, Andari,
  Andeen, Anders, Anders, Anderson, Andreazza, Andrei, Andrieux, Anduaga,
  Angelidakis, Anger, Angerami, Anghinolfi, Anisenkov, Anjos, Annovi, Antonaki,
  Antonelli, Antonov, Antos, Anulli, Aoki, Aoun, Aperio~Bella, Apolle,
  Arabidze, Aracena, Arai, Arce, Arfaoui, Arguin, Arik, Arik, Armbruster,
  Arnaez, Arnal, Arnault, Artamonov, Artoni, Arutinov, Asai, Ask, {\AA}sman,
  Asquith, Assamagan, Astbury, Atkinson, Aubert, Auge, Augsten, Aurousseau,
  Avolio, Avramidou, Axen, Azuelos, Azuma, Baak, Baccaglioni, Bacci, Bach,
  Bachacou, Bachas, Backes, Backhaus, Backus~Mayes, Badescu, Bagnaia,
  Bahinipati, Bai, Bailey, Bain, Baines, Baker, Baker, Baker, Balek, Banas,
  Banerjee, Banerjee, Banfi, Bangert, Bansal, Bansil, Barak, Baranov,
  Barbaro-Galtieri, Barber, Barberio, Barberis, Barbero, Bardin, Barillari,
  Barisonzi, Barklow, Barlow, Barnett, Barnett, Baroncelli, Barone, Barr,
  Barreiro, Barreiro Guimar{\~a}es~da Costa, Barrillon, Bartoldus, Barton,
  Bartsch, Basye, Bates, Batkova, Batley, Battaglia, Battistin, Bauer, Bawa,
  Beale, Beau, Beauchemin, Beccherle, Bechtle, Beck, Becker, Becker,
  Beckingham, Becks, Beddall, Beddall, Bedikian, Bednyakov, Bee, Beemster,
  Begel, Behar~Harpaz, Behera, Beimforde, Belanger-Champagne, Bell, Bell,
  Bella, Bellagamba, Bellomo, Belloni, Beloborodova, Belotskiy, Beltramello,
  Benary, Benchekroun, Bendtz, Benekos, Benhammou, Benhar~Noccioli,
  Benitez~Garcia, Benjamin, Benoit, Bensinger, Benslama, Bentvelsen, Berge,
  Bergeaas~Kuutmann, Berger, Berghaus, Berglund, Beringer, Bernat, Bernhard,
  Bernius, Bernlochner, Berry, Bertella, Bertin, Bertolucci, Besana, Besjes,
  Besson, Bethke, Bhimji, Bianchi, Bianco, Biebel, Bieniek, Bierwagen,
  Biesiada, Biglietti, Bilokon, Bindi, Binet, Bingul, Bini, Biscarat, Bittner,
  Black, Blair, Blanchard, Blanchot, Blazek, Bloch, Blocker, Blocki, Blondel,
  Blum, Blumenschein, Bobbink, Bobrovnikov, Bocchetta, Bocci, Boddy, Boehler,
  Boek, Boelaert, Bogaerts, Bogdanchikov, Bogouch, Bohm, Bohm, Boisvert, Bold,
  Boldea, Bolnet, Bomben, Bona, Boonekamp, Bordoni, Borer, Borisov, Borissov,
  Borjanovic, Borri, Borroni, Bortolotto, Bos, Boscherini, Bosman,
  Boterenbrood, Bouchami, \& Boudrea...}]{2012PhLB..716....1A}
Aad, G., Abajyan, T., Abbott, B., {et~al.} 2012, Phys. Lett. B, 716, 1

\bibitem[{Ackermann {et~al.}(2013)Ackermann, Ajello, Asano, Atwood, Axelsson,
  Baldini, Ballet, Barbiellini, Baring, Bastieri, Bechtol, Bellazzini,
  Bissaldi, Bonamente, Bregeon, Brigida, Bruel, Buehler, Burgess, Buson,
  Caliandro, Cameron, Caraveo, Cecchi, Chaplin, Charles, Chekhtman, Cheung,
  Chiang, Chiaro, Ciprini, Claus, Cleveland, Cohen-Tanugi, Collazzi, Cominsky,
  Connaughton, Conrad, Cutini, D'Ammando, de~Angelis, DeKlotz, de~Palma,
  Dermer, Desiante, Diekmann, Di~Venere, Drell, Drlica-Wagner, Favuzzi, Fegan,
  Ferrara, Finke, Fitzpatrick, Focke, Franckowiak, Fukazawa, Funk, Fusco,
  Gargano, Gehrels, Germani, Gibby, Giglietto, Giles, Giordano, Giroletti,
  Godfrey, Granot, Grenier, Grove, Gruber, Guiriec, Hadasch, Hanabata, Harding,
  Hayashida, Hays, Horan, Hughes, Inoue, Jogler, J{\'o}hannesson, Johnson,
  Kawano, Kn{\"o}dlseder, Kocevski, Kuss, Lande, Larsson, Latronico, Longo,
  Loparco, Lovellette, Lubrano, Mayer, Mazziotta, McEnery, Michelson, Mizuno,
  Moiseev, Monzani, Moretti, Morselli, Moskalenko, Murgia, Nemmen, Nuss, Ohno,
  Ohsugi, Okumura, Omodei, Orienti, Paneque, Pelassa, Perkins, Pesce-Rollins,
  Petrosian, Piron, Pivato, Porter, Racusin, Rain{\`o}, Rando, Razzano,
  Razzaque, Reimer, Reimer, Ritz, Roth, Ryde, Sartori, Parkinson, Scargle,
  Schulz, Sgr{\`o}, Siskind, Sonbas, Spandre, Spinelli, Tajima, Takahashi,
  Thayer, Thayer, Thompson, Tibaldo, Tinivella, Torres, Tosti, Troja, Usher,
  Vandenbroucke, Vasileiou, Vianello, Vitale, Winer, Wood, Yamazaki, Younes,
  Yu, Zhu, Bhat, Briggs, Byrne, Foley, Goldstein, Jenke, Kippen, Kouveliotou,
  McBreen, Meegan, Paciesas, Preece, Rau, Tierney, van~der Horst, von Kienlin,
  Wilson-Hodge, Xiong, Cusumano, La~Parola, \& Cummings}]{Ackermann:2013ih}
Ackermann, M., Ajello, M., Asano, K., {et~al.} 2013, Science, 343, 42

\bibitem[{{Aharonian}(2004)}]{2004vhec.book.....A}
{Aharonian}, F.~A. 2004, {Very high energy cosmic gamma radiation : a crucial
  window on the extreme universe}

\bibitem[{Bernardini(2004)}]{Bernardini:2004cl}
Bernardini, C. 2004, Physics in Perspective, 6, 156

\bibitem[{Bloom {et~al.}(2009)Bloom, Perley, Li, Butler, Miller, Kocevski,
  Kann, Foley, Chen, Filippenko, Starr, Macomber, Prochaska, Chornock,
  Poznanski, Klose, Skrutskie, Lopez, Hall, Glazebrook, \&
  Blake}]{2009ApJ...691..723B}
Bloom, J.~S., Perley, D.~A., Li, W., {et~al.} 2009, ApJ, 691, 723

\bibitem[{Bondi(1952)}]{1952MNRAS.112..195B}
Bondi, H. 1952, MNRAS, 112, 195

\bibitem[{Bondi \& Hoyle(1944)}]{Bondi:1944ty}
Bondi, H., \& Hoyle, F. 1944, MNRAS, 104, 273

\bibitem[{Broderick(2005)}]{2005MNRAS.361..955B}
Broderick, A.~E. 2005, MNRAS, 361, 955

\bibitem[{Cobb {et~al.}(2006)Cobb, Bailyn, van Dokkum, \&
  Natarajan}]{2006ApJ...651L..85C}
Cobb, B.~E., Bailyn, C.~D., van Dokkum, P.~G., \& Natarajan, P. 2006, ApJ, 651,
  L85

\bibitem[{{de Ugarte Postigo} {et~al.}(2013){de Ugarte Postigo}, {Xu},
  {Leloudas}, {Kruehler}, {Malesani}, {Gorosabel}, {Thoene}, {Sanchez-Ramirez},
  {Schulze}, {Fynbo}, {Hjorth}, {Jakobsson}, \&
  {Cabrera-Lavers}}]{2013GCN..14646...1D}
{de Ugarte Postigo}, A., {Xu}, D., {Leloudas}, G., {et~al.} 2013, GCN Circ.,
  14646

\bibitem[{Evans {et~al.}(2007)Evans, Beardmore, Page, Tyler, Osborne, Goad,
  O'Brien, Vetere, Racusin, Morris, Burrows, Capalbi, Perri, Gehrels, \&
  Romano}]{Evans:2007iz}
Evans, P.~A., Beardmore, A.~P., Page, K.~L., {et~al.} 2007, A\&A, 469, 379

\bibitem[{Evans {et~al.}(2009)Evans, Beardmore, Page, Osborne, O'Brien,
  Willingale, Starling, Burrows, Godet, Vetere, Racusin, Goad, Wiersema,
  Angelini, Capalbi, Chincarini, Gehrels, Kennea, Margutti, Morris, Mountford,
  Pagani, Perri, Romano, \& Tanvir}]{Evans:2009kx}
---. 2009, MNRAS, 397, 1177

\bibitem[{Fermi(1934)}]{Fermi:1934}
Fermi, E. 1934, Il Nuovo Cimento, 11, 1

\bibitem[{Fermi(1949)}]{Fermi:1949jk}
---. 1949, Phys. Rev., 75, 1169

\bibitem[{{Flores} {et~al.}(2013){Flores}, {Covino}, {Xu}, {Kruehler}, {Fynbo},
  {Milvang-Jensen}, {de Ugarte Postigo}, {Kaper}, \&
  {Wiersema}}]{2013GCN..14491...1F}
{Flores}, H., {Covino}, S., {Xu}, D., {et~al.} 2013, GCN Circ., 14491

\bibitem[{Friis \& Watson(2013)}]{Friis:2013ef}
Friis, M., \& Watson, D. 2013, ApJ, 771, 15

\bibitem[{Fryer {et~al.}(1996)Fryer, Benz, \& Herant}]{1996ApJ...460..801F}
Fryer, C.~L., Benz, W., \& Herant, M. 1996, ApJ, 460, 801

\bibitem[{Fryer {et~al.}(2014)Fryer, Rueda, \& Ruffini}]{2014arXiv1409.1473F}
Fryer, C.~L., Rueda, J.~A., \& Ruffini, R. 2014, arXiv.org, 1473

\bibitem[{Fryer {et~al.}(1999)Fryer, Woosley, \&
  Hartmann}]{1999ApJ...526..152F}
Fryer, C.~L., Woosley, S.~E., \& Hartmann, D.~H. 1999, ApJ, 526, 152

\bibitem[{Galama {et~al.}(1998)Galama, Vreeswijk, van Paradijs, Kouveliotou,
  Augusteijn, B{\"o}hnhardt, Brewer, Doublier, Gonzalez, Leibundgut, Lidman,
  Hainaut, Patat, Heise, in't Zand, Hurley, Groot, Strom, Mazzali, Iwamoto,
  Nomoto, Umeda, Nakamura, Young, Suzuki, Shigeyama, Koshut, Kippen, Robinson,
  de~Wildt, Wijers, Tanvir, Greiner, Pian, Palazzi, Frontera, Masetti,
  Nicastro, Feroci, Costa, Piro, Peterson, Tinney, Boyle, Cannon, Stathakis,
  Sadler, Begam, \& Ianna}]{Galama:1998ea}
Galama, T.~J., Vreeswijk, P.~M., van Paradijs, J., {et~al.} 1998, Nature, 395,
  670

\bibitem[{Giacconi \& Ruffini(1975)}]{Giacconi1975}
Giacconi, R., \& Ruffini, R. 1975, {Physics and Astrophysics of Neutron Stars
  and Black Holes}

\bibitem[{Grupe {et~al.}(2007)Grupe, Gronwall, Wang, Roming, Cummings, Zhang,
  Meszaros, Trigo, O'Brien, Page, Beardmore, Godet, vanden Berk, Brown, Koch,
  Morris, Stroh, Burrows, Nousek, McMath~Chester, Immler, Mangano, Romano,
  Chincarini, Osborne, Sakamoto, \& Gehrels}]{2007ApJ...662..443G}
Grupe, D., Gronwall, C., Wang, X.-Y., {et~al.} 2007, ApJ, 662, 443

\bibitem[{Guetta \& Della~Valle(2007)}]{Guetta:2007tb}
Guetta, D., \& Della~Valle, M. 2007, ApJL, 657, L73

\bibitem[{Gupta \& Zhang(2008)}]{Gupta:2008cc}
Gupta, N., \& Zhang, B. 2008, MNRAS, 384, L11

\bibitem[{Izzo {et~al.}(2012)Izzo, Rueda, \& Ruffini}]{2012A&A...548L...5I}
Izzo, L., Rueda, J.~A., \& Ruffini, R. 2012, A\&A, 548, L5

\bibitem[{Kouveliotou {et~al.}(2013)Kouveliotou, Granot, Racusin, Bellm,
  Vianello, Oates, Fryer, Boggs, Christensen, Craig, Dermer, Gehrels, Hailey,
  Harrison, Melandri, McEnery, Mundell, Stern, Tagliaferri, \&
  Zhang}]{Kouveliotou:2013jx}
Kouveliotou, C., Granot, J., Racusin, J.~L., {et~al.} 2013, ApJ, 779, L1

\bibitem[{Kovacevic {et~al.}(2014)Kovacevic, Izzo, Wang, Muccino, Della~Valle,
  Amati, Barbarino, Enderli, Pisani, \& Li}]{Kovacevic:2014up}
Kovacevic, M., Izzo, L., Wang, Y., {et~al.} 2014, 1408.6227

\bibitem[{Laskar {et~al.}(2013)Laskar, Berger, Zauderer, Margutti, Soderberg,
  Chakraborti, Lunnan, Chornock, Chandra, \& Ray}]{2013ApJ...776..119L}
Laskar, T., Berger, E., Zauderer, B.~A., {et~al.} 2013, ApJ, 776, 119

\bibitem[{{Levan} {et~al.}(2013){Levan}, {Cenko}, {Perley}, \&
  {Tanvir}}]{2013GCN..14455...1L}
{Levan}, A.~J., {Cenko}, S.~B., {Perley}, D.~A., \& {Tanvir}, N.~R. 2013, GCN
  Circ., 14455

\bibitem[{Levan {et~al.}(2013{\natexlab{a}})Levan, Fruchter, Graham, Tanvir,
  Hjorth, Fynbo, Perley, Cenko, Pian, Cano, Pe'Er, Hounsell, Mishra, \&
  Kouveliotou}]{2013GCN..14686...1L}
Levan, A.~J., Fruchter, A.~S., Graham, J., {et~al.} 2013{\natexlab{a}}, GCN
  Circ., 1468

\bibitem[{Levan {et~al.}(2013{\natexlab{b}})Levan, Tanvir, Fruchter, Hjorth,
  Pian, Mazzali, Perley, Cano, Graham, Hounsell, Cenko, Fynbo, Kouveliotou,
  Pe'er, Misra, \& Wiersema}]{2013arXiv1307.5338L}
Levan, A.~J., Tanvir, N.~R., Fruchter, A.~S., {et~al.} 2013{\natexlab{b}},
  arXiv, 5338

\bibitem[{Li \& Paczynski(1998)}]{1998ApJ...507L..59L}
Li, L.-X., \& Paczynski, B. 1998, ApJ, 507, L59

\bibitem[{Lithwick \& Sari(2001)}]{Lithwick:2001vg}
Lithwick, Y., \& Sari, R. 2001, ApJ, 555, 540

\bibitem[{{Maselli} {et~al.}(2013){Maselli}, {Beardmore}, {Lien}, {Mangano},
  {Mountford}, {Page}, {Palmer}, \& {Siegel}}]{2013GCN..14448...1M}
{Maselli}, A., {Beardmore}, A.~P., {Lien}, A.~Y., {et~al.} 2013, GCN Circ.,
  14448

\bibitem[{Maselli {et~al.}(2013)Maselli, Melandri, Nava, Mundell, Kawai,
  Campana, Covino, Cummings, Cusumano, Evans, Ghirlanda, Ghisellini, Guidorzi,
  Kobayashi, Kuin, La~Parola, Mangano, Oates, Sakamoto, Serino, Virgili, Zhang,
  Barthelmy, Beardmore, Bernardini, Bersier, Burrows, Calderone, Capalbi,
  Chiang, D'Avanzo, D'Elia, de~Pasquale, Fugazza, Gehrels, Gomboc, Harrison,
  Hanayama, Japelj, Kennea, Kopa{\v c}, Kouveliotou, Kuroda, Levan, Malesani,
  Marshall, Nousek, O'Brien, Osborne, Pagani, Page, Page, Perri, Pritchard,
  Romano, Saito, Sbarufatti, Salvaterra, Steele, Tanvir, Vianello, Weigand,
  Wiersema, Yatsu, Yoshii, \& Tagliaferri}]{Maselli:2013hc}
Maselli, A., Melandri, A., Nava, L., {et~al.} 2013, Science

\bibitem[{M{\'e}sz{\'a}ros(2006)}]{Meszaros:2006gn}
M{\'e}sz{\'a}ros, P. 2006, Rep. Prog. Phys., 69, 2259

\bibitem[{Muccino {et~al.}(2013{\natexlab{a}})Muccino, Ruffini, Bianco, Izzo,
  \& Penacchioni}]{Muccino:2013cq}
Muccino, M., Ruffini, R., Bianco, C.~L., Izzo, L., \& Penacchioni, A.~V.
  2013{\natexlab{a}}, ApJ, 763, 125

\bibitem[{Muccino {et~al.}(2013{\natexlab{b}})Muccino, Ruffini, Bianco, Izzo,
  Penacchioni, \& Pisani}]{Muccino:2013et}
Muccino, M., Ruffini, R., Bianco, C.~L., {et~al.} 2013{\natexlab{b}}, ApJ, 772,
  62

\bibitem[{Muccino {et~al.}(2014)Muccino, Bianco, Izzo, Wang, Enderli,
  Kovacevic, Pisani, Penacchioni, \& Ruffini}]{Muccino:2014ex}
Muccino, M., Bianco, C.~L., Izzo, L., {et~al.} 2014, Gravitation and Cosmology,
  20, 197

\bibitem[{Nava {et~al.}(2013)Nava, Sironi, Ghisellini, Celotti, \&
  Ghirlanda}]{Nava:2013gf}
Nava, L., Sironi, L., Ghisellini, G., Celotti, A., \& Ghirlanda, G. 2013,
  MNRAS, 433, 2107

\bibitem[{{Nousek} {et~al.}(2006){Nousek}, {Kouveliotou}, {Grupe}, {Page},
  {Granot}, {Ramirez-Ruiz}, {Patel}, {Burrows}, {Mangano}, {Barthelmy},
  {Beardmore}, {Campana}, {Capalbi}, {Chincarini}, {Cusumano}, {Falcone},
  {Gehrels}, {Giommi}, {Goad}, {Godet}, {Hurkett}, {Kennea}, {Moretti},
  {O'Brien}, {Osborne}, {Romano}, {Tagliaferri}, \&
  {Wells}}]{2006ApJ...642..389N}
{Nousek}, J.~A., {Kouveliotou}, C., {Grupe}, D., {et~al.} 2006, ApJ, 642, 389

\bibitem[{Page {et~al.}(2011)Page, Starling, Fitzpatrick, Pandey, Osborne,
  Schady, McBreen, Campana, Ukwatta, Pagani, Beardmore, \&
  Evans}]{2011MNRAS.416.2078P}
Page, K.~L., Starling, R. L.~C., Fitzpatrick, G., {et~al.} 2011, MNRAS, 416,
  2078

\bibitem[{{Penacchioni} {et~al.}(2013){Penacchioni}, {Ruffini}, {Bianco},
  {Izzo}, {Muccino}, {Pisani}, \& {Rueda}}]{2013A&A...551A.133P}
{Penacchioni}, A.~V., {Ruffini}, R., {Bianco}, C.~L., {et~al.} 2013, A\&A, 551,
  A133

\bibitem[{Penacchioni {et~al.}(2012)Penacchioni, Ruffini, Izzo, Muccino,
  Bianco, Caito, Patricelli, \& Amati}]{2012A&A...538A..58P}
Penacchioni, A.~V., Ruffini, R., Izzo, L., {et~al.} 2012, A\&A, 538, 58

\bibitem[{Perley {et~al.}(2013)Perley, Cenko, Corsi, Tanvir, Levan, Kann,
  Sonbas, Wiersema, Zheng, Zhao, Bai, Chang, Clubb, Frail, Fruchter, Greiner,
  G{\"u}ver, Horesh, Filippenko, Klose, Mao, Morgan, Schmidl, Stecklum, Tanga,
  Wang, \& Xin}]{Perley:2013tf}
Perley, D.~A., Cenko, S.~B., Corsi, A., {et~al.} 2013, ApJ, 781, 37

\bibitem[{Pian {et~al.}(2000)Pian, Amati, Antonelli, Butler, Costa, Cusumano,
  Danziger, Feroci, Fiore, Frontera, Giommi, Masetti, Muller, Nicastro,
  Oosterbroek, Orlandini, Owens, Palazzi, Parmar, Piro, in~t Zand,
  Castro~Tirado, Coletta, Dal~Fiume, Del~Sordo, Heise, Soffitta, \&
  Torroni}]{Pian:2000dd}
Pian, E., Amati, L., Antonelli, L.~A., {et~al.} 2000, ApJ, 536, 778

\bibitem[{Piran(2005)}]{Piran:2005cs}
Piran, T. 2005, RMP, 76, 1143

\bibitem[{Pisani {et~al.}(2013)Pisani, Izzo, Ruffini, Bianco, Muccino,
  Penacchioni, Rueda, \& Wang}]{Pisani:2013id}
Pisani, G.~B., Izzo, L., Ruffini, R., {et~al.} 2013, A\&A, 552, L5

\bibitem[{Rhoads(1999)}]{1999ApJ...525..737R}
Rhoads, J.~E. 1999, ApJ, 525, 737

\bibitem[{Romano {et~al.}(2006)Romano, Campana, Chincarini, Cummings, Cusumano,
  Holland, Mangano, Mineo, Page, Pal'shin, Rol, Sakamoto, Zhang, Aptekar,
  Barbier, Barthelmy, Beardmore, Boyd, Burrows, Capalbi, Fenimore, Frederiks,
  Gehrels, Giommi, Goad, Godet, Golenetskii, Guetta, Kennea, La~Parola,
  Malesani, Marshall, Moretti, Nousek, O'Brien, Osborne, Perri, \&
  Tagliaferri}]{Romano:2006kt}
Romano, P., Campana, S., Chincarini, G., {et~al.} 2006, A\&A, 456, 917

\bibitem[{Rueda \& Ruffini(2012)}]{2012ApJ...758L...7R}
Rueda, J.~A., \& Ruffini, R. 2012, ApJL, 758, L7

\bibitem[{Ruffini {et~al.}(2001)Ruffini, Bianco, Fraschetti, Xue, \&
  Chardonnet}]{2001ApJ...555L.117R}
Ruffini, R., Bianco, C.~L., Fraschetti, F., Xue, S.-S., \& Chardonnet, P. 2001,
  ApJ, 555, L117

\bibitem[{Ruffini {et~al.}(1999)Ruffini, Salmonson, Wilson, \&
  Xue}]{1999A&A...350..334R}
Ruffini, R., Salmonson, J.~D., Wilson, J.~R., \& Xue, S.-S. 1999, Astronomy and
  Astrophysics, 350, 334

\bibitem[{Ruffini {et~al.}(2000)Ruffini, Salmonson, Wilson, \&
  Xue}]{2000A&A...359..855R}
---. 2000, Astronomy and Astrophysics, 359, 855

\bibitem[{Ruffini {et~al.}(2014)Ruffini, Vereshchagin, \& Wang}]{Ruffini2014c}
Ruffini, R., Vereshchagin, V., \& Wang, Y. 2014, In preparation

\bibitem[{{Ruffini} \& {Wilson}(1975)}]{1975PhRvD..12.2959R}
{Ruffini}, R., \& {Wilson}, J.~R. 1975, PRD, 12, 2959

\bibitem[{Ruffini {et~al.}(2008)Ruffini, Bernardini, Bianco, Caito, Chardonnet,
  Cherubini, Dainotti, Fraschetti, Geralico, Guida, Patricelli, Rotondo,
  Rueda~Hernandez, Vereshchagin, \& Xue}]{2008mgm..conf..368R}
Ruffini, R., Bernardini, M.~G., Bianco, C.~L., {et~al.} 2008, Proceeding of The
  Eleventh Marcel Grossmann Meeting, 368

\bibitem[{{Ruffini} {et~al.}(2013{\natexlab{a}}){Ruffini}, {Bianco}, {Enderli},
  {Kovacevic}, {Muccino}, {Penacchioni}, {Pisani}, {Rueda}, \&
  {Wang}}]{RuffiniGCN1}
{Ruffini}, R., {Bianco}, C.~L., {Enderli}, M., {et~al.} 2013{\natexlab{a}}, GCN
  Circ., 15560, 1

\bibitem[{{Ruffini} {et~al.}(2013{\natexlab{b}}){Ruffini}, {Bianco}, {Enderli},
  {Muccino}, {Penacchioni}, {Pisani}, {Rueda}, {Sahakyan}, {Wang}, \&
  {Izzo}}]{2013GCN..14526...1R}
---. 2013{\natexlab{b}}, GCN Circ., 14526

\bibitem[{{Ruffini} {et~al.}(2013{\natexlab{c}}){Ruffini}, {Bianco}, {Enderli},
  {Kovacevic}, {Muccino}, {Penacchioni}, {Pisani}, {Rueda}, \&
  {Wang}}]{RuffiniGCN2}
---. 2013{\natexlab{c}}, GCN Circ., 15576, 1

\bibitem[{{Ruffini} {et~al.}(2014{\natexlab{a}}){Ruffini}, {Bianco}, {Enderli},
  {Kovacevic}, {Muccino}, {Penacchioni}, {Pisani}, {Rueda}, \&
  {Wang}}]{RuffiniGCN3}
---. 2014{\natexlab{a}}, GCN Circ., 15794, 1

\bibitem[{{Ruffini} {et~al.}(2014{\natexlab{b}}){Ruffini}, {Muccino}, {Bianco},
  {Enderli}, {Izzo}, {Kovacevic}, {Penacchioni}, {Pisani}, {Rueda}, \&
  {Wang}}]{2014A&A...565L..10R}
{Ruffini}, R., {Muccino}, M., {Bianco}, C.~L., {et~al.} 2014{\natexlab{b}},
  A\&A, 565, L10

\bibitem[{Ruffini {et~al.}(2014)Ruffini, Izzo, Muccino, Pisani, Rueda, Wang,
  Barbarino, Bianco, Enderli, \& Kovacevic}]{2014arXiv1404.1840R}
Ruffini, R., Izzo, L., Muccino, M., {et~al.} 2014, arXiv, 1840

\bibitem[{Starling {et~al.}(2012)Starling, Page, Pe'er, Beardmore, \&
  Osborne}]{2012MNRAS.427.2950S}
Starling, R. L.~C., Page, K.~L., Pe'er, A., Beardmore, A.~P., \& Osborne, J.~P.
  2012, MNRAS, 427, 2950

\bibitem[{Tanvir {et~al.}(2010)Tanvir, Rol, Levan, Svensson, Fruchter, Granot,
  O'Brien, Wiersema, Starling, Jakobsson, Fynbo, Hjorth, Curran, van~der Horst,
  Kouveliotou, Racusin, Burrows, \& Genet}]{2010ApJ...725..625T}
Tanvir, N.~R., Rol, E., Levan, A.~J., {et~al.} 2010, ApJ, 725, 625

\bibitem[{{Tavani}(1998)}]{1998ApJ...497L..21T}
{Tavani}, M. 1998, ApJL, 497, L21

\bibitem[{Trotter {et~al.}(2013)Trotter, Reichart, Haislip, Lacluyze, McLin,
  Cominsky, Smith, Caton, Hawkins, Holmes, Linder, Berger, Cromartie, Egger,
  Foster, Frank, Ivarsen, Maples, Moore, Nysewander, Speckhard, \&
  Crain}]{2013GCN..14662...1T}
Trotter, A., Reichart, D., Haislip, J., {et~al.} 2013, GCN Circ., 1466

\bibitem[{van Eerten {et~al.}(2010)van Eerten, Zhang, \&
  MacFadyen}]{vanEerten:2010gw}
van Eerten, H., Zhang, W., \& MacFadyen, A. 2010, ApJ, 722, 235

\bibitem[{van Eerten \& MacFadyen(2012)}]{vanEerten:2012hp}
van Eerten, H.~J., \& MacFadyen, A.~I. 2012, ApJ, 751, 155

\bibitem[{{von Kienlin}(2013)}]{2013GCN..14473...1V}
{von Kienlin}, A. 2013, GCN Circ., 14473

\bibitem[{{Woosley}(1993)}]{1993ApJ...405..273W}
{Woosley}, S.~E. 1993, ApJ, 405, 273

\bibitem[{{Xu} {et~al.}(2013){Xu}, {de Ugarte Postigo}, {Schulze},
  {Jessen-Hansen}, {Leloudas}, {Kruehler}, {Fynbo}, \&
  {Jakobsson}}]{2013GCN..14478...1X}
{Xu}, D., {de Ugarte Postigo}, A., {Schulze}, S., {et~al.} 2013, GCN Circ.,
  14478

\bibitem[{Xu {et~al.}(2013)Xu, de~Ugarte~Postigo, Leloudas, Kr{\"u}hler, Cano,
  Hjorth, Malesani, Fynbo, Th{\"o}ne, S{\'a}nchez-Ram{\'\i}rez, Schulze,
  Jakobsson, Kaper, Sollerman, Watson, Cabrera-Lavers, Cao, Covino, Flores,
  Geier, Gorosabel, Hu, Milvang-Jensen, Sparre, Xin, Zhang, Zheng, \&
  Zou}]{Xu:2013ic}
Xu, D., de~Ugarte~Postigo, A., Leloudas, G., {et~al.} 2013, ApJ, 776, 98

\bibitem[{Zel'dovich {et~al.}(1972)Zel'dovich, Ivanova, \&
  Nadezhin}]{1972SvA....16..209Z}
Zel'dovich, Y.~B., Ivanova, L.~N., \& Nadezhin, D.~K. 1972, Soviet Astron., 16,
  209

\end{thebibliography}
\end{document}